\documentclass[aps,superscriptaddress,prr,twocolumn]{revtex4-2}

\usepackage{amsmath}
\usepackage{amssymb}
\usepackage{graphicx}
\usepackage{color}
\usepackage{bm}
\usepackage{mathrsfs}
\usepackage{multirow}
\usepackage{times}

\def\rr#1{(\ref{#1})}

\newcommand{\be}{\begin{equation}}
\newcommand{\ee}{\end{equation}}

\newcommand{\appsection}[1]{\section{\MakeUppercase{#1}}} 

\begin{document}
\title{Self-generated electrokinetic flows from active-charged boundary patterns}

\author{Ahis Shrestha}

\affiliation{Center for Computation and Theory of Soft Materials, Northwestern University, Evanston, IL 60208}

\affiliation{Department of Physics and Astronomy, Northwestern University, Evanston, IL 60208}

\author{Eleftherios Kirkinis}
\affiliation{Center for Computation and Theory of Soft Materials, Northwestern University, Evanston, IL 60208}
\affiliation{Department of Material Science and Engineering, Northwestern University, Evanston, IL 60208}

\author{Monica Olvera de la Cruz}

\affiliation{Center for Computation and Theory of Soft Materials, Northwestern University, Evanston, IL 60208}

\affiliation{Department of Material Science and Engineering, Northwestern University, Evanston, IL 60208}
%\affiliation{Department of Chemistry, Northwestern University, Evanston, IL 60208}
\affiliation{Department of Physics and Astronomy, Northwestern University, Evanston, IL 60208}

%\date{\today}

\begin{abstract}

We develop a hydrodynamic description of self-generated electrolyte flow in capillaries whose bounding walls feature \emph{both} non-uniform distributions of charge \emph{and} non-uniform active ionic fluxes. The hydrodynamic velocity arising in such a system has components that are forbidden by symmetry in the absence of charge and fluxes. However, when these two boundary mechanisms are simultaneously present, they can lead to a symmetry broken state where steady flows with both unidirectional and circulatory components emerge. We show that these flow states arise when modulated boundary patterns of charge and fluxes are offset by a flux-charge phase difference, which is associated with the separation between sites of their peak densities on the wall. Mismatch in diffusivity of cationic and anionic species can modify the flow states and becomes an enhancing factor when fluxes of both ion species are being produced together at the same site. We demonstrate that this mechanism can be realized with a microfluidic generator which is powered by enzyme-coated patches that catalyzes reactants in the solution to produce fluxes of ions. The local ionic elevation or depletion that disrupts a non-uniform double layer, promotes self-induced gradients yielding persistent body forces to generate bulk fluid motion. Our work quantifies a boundary-driven mechanism behind self-sustained electrolyte flow in confined environments that exists without any external bulk-imposed fields or gradients. It provides a theoretical framework for understanding the combined effect of active and charged boundaries that are relevant in biological or soft matter systems, and can be utilized in electrofluidic and iontronic applications.   

\end{abstract}

\maketitle

\section{Introduction} \label{secIntro}
Self-sustained fluid motion is an essential mechanism involved in various cellular processes that regulate intracellular transport \cite{Woodhouse2013, Goldstein2015, Monteith2016, Halatek2018, Htet2023}. Micro- and nano-fluidic technologies inspired by biological or soft matter systems have witnessed rapid development in recent years. This includes the emerging field of iontronics which employs ions as charge carriers, for instance, in neuromorphic devices that can be comprised of fluidic chambers analogous to the axon of neurons \cite{Robin2021,Robin2023,Kamsma2023a,Kamsma2023b, Harikesh2022}. The directed flow of charged fluids such as an ionic solution or electrolyte in confined environments has thus become an important area of investigation. In particular, the underlying physical parameters and confinement boundary properties that control and mediate such flows need to be understood and clarified \cite{Kavokine2021, JimenezAngeles2023, Qu2021,Woodhouse2017,Henrique2024,Wang2024}.  

A well-known mechanism for micro- and nano-scale transport of electrolyte near charged interfaces employs gradients of electric potential or ionic concentration. These are types of electrokinetic flows referred to as electro-osmosis or ionic diffusio-osmosis, respectively. On one hand, this can be achieved through externally applied electric fields or by imposing bulk ionic gradients. Mobile ions and fixed surface charges are brought into balance by forming an electric double layer (EDL) at the interface, and the applied field or gradient exerts body forces acting within this layer where the ions are dislocated along with the surrounding viscous liquid, causing a net fluid motion in response \cite{Anderson1989,Dukhin1993,Lyklema1995,Kirby2010,Dominguez2022}. On the other hand, electrokinetic flows can alternatively be induced through surface-generated ionic fluxes that are spatially nonuniform or asymmetric \cite{Paxton2006,Ristenpart2008,Davidson2018,Ashaju2021}. Variations in fluxes along the surface lead to local ionic gradients, which in turn induce fluid flow. Ionic fluxes can naturally occur at active sites such as membrane ion pumps in a biological context \cite{Hille2001,Siegel2011,Toko1985, Babourina2004}, or can be chemically produced via surface reactants grafted in catalytic micro-channels or capillaries \cite{Sengupta2014,Gao2022,Wang2023,Song2024,Popescu2025}. Furthermore, in the realm of self-propelling active colloids, it has been shown that charged Janus particles with asymmetric fluxes of ions on their surface move via ionic self-phoresis \cite{Illien2017,Zhou2018,DeCorato2020,Shrestha2024,Dominguez2024}. 

Spatial variations of interfacial properties such as surface charge are also known to affect electrokinetic flow fields and net velocity \cite{Ober2021,Malgaretti2014,Antunes2022}. In the case of externally driven elecroosmosis, both theoretical  \cite{Anderson1985,Ajdari1995,Ajdari1996,Long1999} and experimental \cite{Stroock2000,Stone2004,Biddiss2004,Paratore2019} works have shown that surface charge variation leads to different flow patterns. Heterogeneous surface charge triggers additional gradients on top of the constant applied field, and induce nonuniform flows in the bulk such as recirculating and multidirectional flows when alternating charges are patterned periodically on channel walls. This feature can be utilized in fluidic devices that are functionalized for mixing and pumping of ionic solutions.

\begin{figure}[t!]
    \centering
    \includegraphics[width=1\columnwidth]{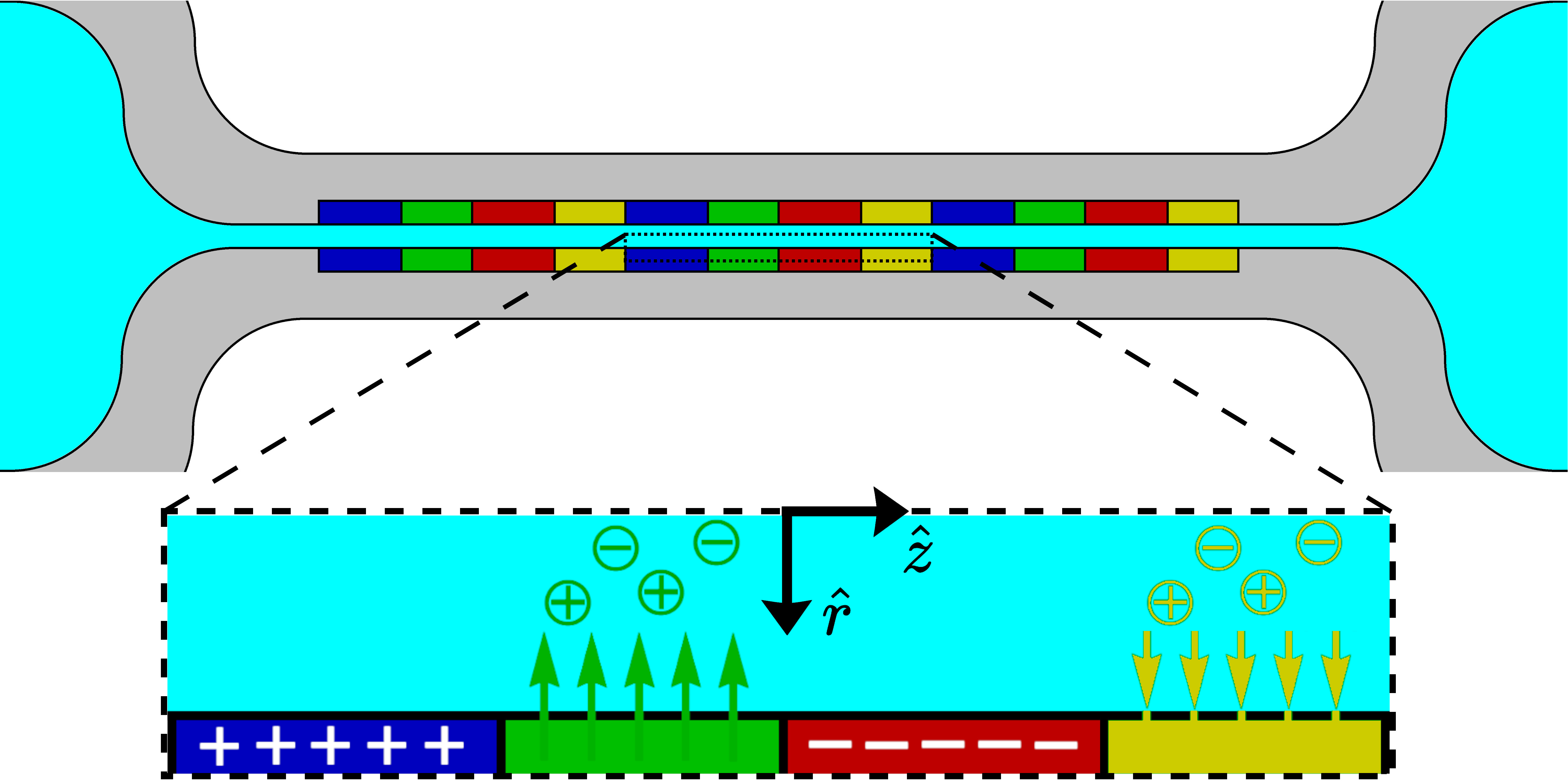}
    \caption{Illustration of a capillary with periodically patterned active-charged surfaces. Top: the cross section of a long cylindrical capillary connecting two identical electrolyte reservoirs. Bottom: the zoomed-in view of the interface between the patterned capillary walls and the electrolyte inside. Shown here are charged regions with positive (blue) and negative (red) surface charge, and active regions with ($+$,$-$) ionic fluxes inwards (green) and outwards (yellow). We use axisymmetric cylindrical coordinates with radial $\hat{r}$ and axial $\hat{z}$ unit vectors as indicated.}
    \label{fig1}
\end{figure}

\begin{figure*}[t!]
    \centering
    \includegraphics[width=0.95\textwidth]{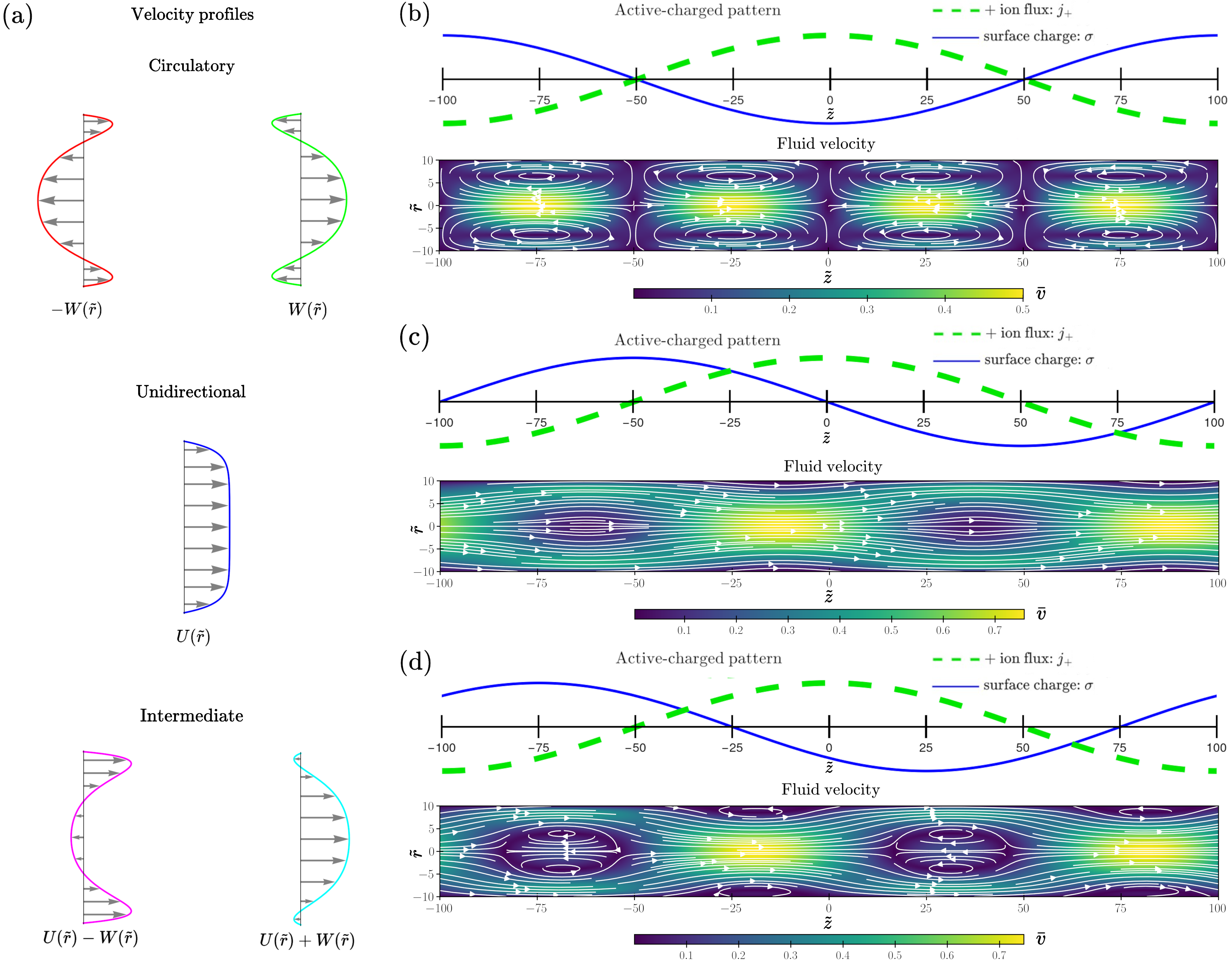}
    \caption{Flow states produced by flux and charge boundary modulations. (a) The $z$-component velocity profile shapes for circulatory $\pm W$,  unidirectional $U$, and intermediate $U \pm W$. The stream plots of steady state fluid velocity in $(\tilde{r},\tilde{z})$ obtained analytically from Eq.~(\ref{eqVr}) and~(\ref{eqVz}) discussed in Sec.~\ref{secFlow} for case of zero anionic flux $j_{0-}=0$ and same relative flux-charge peak amplitudes $\delta_+=\delta_\sigma$ in Eq.~(\ref{deltas}), at different flux-charge phase difference $\beta$ (with $\alpha_+=0$): leading to (b) fully circulatory flow at $\beta=\pi$ and (c) unidirectional dominant flow at $\beta=\pi/2$, and (d) intermediate flow at $\beta=3\pi/4$. The corresponding modulation forms of the cationic (+) flux and the surface charge are shown above. The arrows of streamlines indicate the flow direction, and the color map indicates the velocity magnitude. We show here in dimensionless units of $\tilde{r}=r/\lambda$, $\tilde{z}=z/\lambda$ and  $\bar{v}= v/\lambda$ s$^{-1}$ for a capillary with radius $R=10 \lambda$ and modulation wavelength $l=200 \lambda$ given in terms of the Debye length $\lambda$.}
    \label{fig2}
\end{figure*}

Whereas unidirectional electrolyte flow induced by time-dependent boundary voltages has been realized both experimentally \cite{Green2000,GarciaSanchez2009} and theoretically \cite{Ehrlich1982,Ajdari2000,Cahill2004,Shrestha2025}, a \emph{time-independent} concept that would replace these approaches and circumvent early challenges, is still elusive. In this paper we show that the hydrodynamic velocity of an electrolyte confined by walls endowed with \emph{time-independent} spatially-varying charge \emph{and} ionic activity (Fig.~\ref{fig1}), contains terms which exist only when the aforementioned boundary mechanisms are simultaneously present. The significance of these terms is that they describe a physical effect which is absent in systems lacking these boundary features. We emphasize that this is a boundary-driven effect that exists without any external bulk-imposed electric fields and ionic concentration gradients. In particular, we establish the existence of a zero mode velocity, that is, a non-vanishing unidirectional velocity component that is self-generated in conjunction with a circulatory component along the capillary. Disruption of a non-uniform double layer via local ionic elevation or depletion, leads to strong body forces driving the bulk viscous flow. We find that this effect arises from a flux-charge phase difference associated with spatial separation between sites of peak flux and charge densities on the boundary. Moreover, a mismatch in diffusivity between cationic and anionic species modifies the flow states and becomes an enhancing factor when fluxes of both cations and anions are being produced together at the same site. Estimates of the effect at various electrolyte concentrations and conditions show that it is of considerable magnitude and thus expected be observed with great facility in an experiment. We further demonstrate how this flux-charge mechanism can be realized in a concrete example of a microfluidic generator powered by chemical reactions with enzyme-coated wall patches.   

We employ the continuum framework of the Poisson-Nernst-Planck-Navier-Stokes (PNPNS) equations, and carry out theoretical analytic approximations and finite-element numerical simulations. We show that boundary-modulated charge and active ionic fluxes self-generate two basic flow state components, a circulatory and a unidirectional (zero mode) along the capillary length, as shown in panels (b) and (c) of Fig. \ref{fig2}, respectively. These flow states depend on the phase differences between ionic fluxes and charge modulation patterns and are determined by a number of selection rules. Because the EDL is non-uniform in the longitudinal direction of the capillary, due to the varying wall charge coupled with strong ionic fluxes, there is significant ionic elevation and depletion, constantly knocking it off a uniform equilibrium state. Consequently, this leads to self-induced ionic gradients which are in-phase or out-of phase with respect to the commensurate electric field. In the latter case the electric body force is such as to drive the circulation of a sequence of adjacent vortices. In the former case the electric body force is always of a single sign and drives the electrolyte in a single direction. We also find that the switching of cationic-anionic diffusivity can lead to velocity reversal and change of the flow state.  

The mechanism of flux-charge separation and diffusivity mismatch can be realized with an enzyme powered microfluidic generator \cite{Sengupta2014,Song2024,DeCorato2020,Popescu2025}. In particular, we demonstrate this using a microfluidic chamber as in Fig.~\ref{fig1} where an urease-coated patch \cite{Song2024,DeCorato2020} is grafted in the capillary wall which catalyzes reactants in the liquid solution releasing a localized in-flux of ions as end products. Together with this flux patch, we also consider a separate wall patch constructed with silica \cite{DeCorato2020} or polymer coating \cite{Stroock2000,Biddiss2004} which carries some local surface charge. We show that such a pair of patches generates unidirectional and circulatory flows with localized vortex patterns and the flow speed depends on the flux-charge patch separation distance. Furthermore, we compare two cases of previous works on urease patches \cite{Song2024,DeCorato2020} where the pair of cation-anion end products are different with distinct diffusivity mismatch, and find that the flow speed is enhanced by the larger mismatch. Taken together, our work describes a boundary-driven mechanism based on the combined effect of non-uniform patterning of active and charged sites that controls the transport of confined electrolytes.

This article has been organized as follows. In Sec.~\ref{secSym} we establish the form the velocity components acquire in the presence of charge and flux amplitudes as these are allowed by symmetry. Sec.~\ref{secTheory} describes the continuum PNPNS theory which we employ to obtain the flow patterns in the small-amplitude and long-wavelength approximations. In Sec.~\ref{secFlow} we employ numerical simulations to illustrate and discuss the different flow states generated for special cases of ionic activity and study their dependence on phase difference of the flux and charge modulation. These conclusions are corroborated by the analytical approximations developed in section ~\ref{secTheory}. We quantify and provide estimates of the net unidirectional flow in Sec.~\ref{secAvg} showing that it can reach considerable magnitudes. Likewise, in sec.~\ref{secEffect} we employ analytical approximations and numerical simulations to demonstrate the effect of ionic diffusivity mismatch. Then we present in Sec.~\ref{secExample} a concrete example of an enzyme powered microfluidic generator. Finally, we conclude by summarizing our main results and giving an outlook on future directions in Sec.~\ref{secConc}. The Appendices include details of our analytical calculations and details of the finite-element numerical techniques. 

\section{Symmetry of self-generated flow} \label{secSym}
Simple symmetry arguments can lead to a prediction of the form of the velocity field in the self-generated electrokinetic flow developed in this paper. 
Consider a long cylindrical capillary of radius $R$ containing a 1:1 electrolyte consisting of a cationic ($+$) and anionic ($-$) species which are both monovalent with valencies $Z_{\pm}=\pm 1$ and carrying elementary charge $e$. The capillary connects two identical electrolyte reservoirs that are electroneutral at the bulk where the concentration (number density) of the ion species are equivalent. The capillary walls are patterned with a space-periodic distribution of the ionic flux and charge density (see Fig.~\ref{fig1}). Here we focus on examining the role of surface patterning through phase differences between the modulation modes of ionic fluxes and charge variations. In particular, we take the (surface) flux of the cations and anions at the capillary wall to be of the form 
\begin{equation} \label{wjpm} 
      j_\pm(z) = j_{0 \pm}  \cos{(k z+\alpha_\pm)}   \quad \textrm{at}\quad  r=R
\end{equation}
respectively, and the surface charge to be
\begin{equation} \label{ws} 
      \sigma(z) = \sigma_0  \cos{(k z + \beta)}   \quad \textrm{at}\quad  r=R
\end{equation}
where $k=2\pi/l$ is the modulation wavenumber, $l$ its wavelength and $\alpha_\pm$ and $\beta$ are respective phases. 
Thus, the conditions satisfied by the self-generated electric field and by the net volume flux at the wall are 
$-\varepsilon \mathbf{E} \cdot \hat{r} = \sigma(z)$ and 
 $- \mathbf{J}_\pm \cdot \hat{r} = j_\pm (z)$, respectively, $\hat{r}$ is the radial unit vector, $\varepsilon$ is the dielectric constant and the definition of the net volume flux $ \mathbf{J}_\pm$ follows up in section \ref{secTheory}.  
The ionic fluxes and surface charge carry peak amplitudes $j_{0+}$, $j_{0-}$ and $\sigma_0$. For simplicity we take the patterning to be axisymmetric. 

The hydrodynamic velocity of the electrolyte arising in such a confinement has radial and axial components described by
$
\mathbf{v} = v_r(r,z) \hat{{r}} + v_z(r,z) \hat{{z}}
$
in the axisymmetric cylindrical coordinates $(r,z)$ as these are displayed in figure \ref{fig1}. In the presence of both boundary effects, the following contributions to the axial component are allowed
\begin{equation} \label{sym}
v_z = \gamma_{m,m'}j_{0 m} j_{0 m'} +  \gamma_{m,\sigma}j_{0 m} \sigma_0 
\end{equation}
where $m,m' \in \{+,-\}$ and summation is implied over repeated indices. The coefficients $\gamma$ above are \emph{odd} functions of $k$. Since the velocity is a polar vector, it changes sign under spatial and domain inversion $(r,z) \rightarrow(-r,-z)$ and $k \rightarrow -k$. We show below that the first term in the right hand side of \rr{sym} gives rise to circulatory behavior and the second term leads a zero mode which unidirectionally drives the electrolyte along the capillary, for example shown in panels (b) and (c) of Fig. \ref{fig2}, respectively.

\section{Continuum hydrodynamic theory} \label{secTheory}

The characteristic length scale associated with the size of the EDL is the Debye length given by $\lambda=\sqrt{\varepsilon k_BT/e^2C_\infty}$, where $\varepsilon$ is the electric permittivity and $k_BT$ is the characteristic thermal energy of liquid water at room temperature. Here $C_\infty$ is the total bulk ion concentration such that both ionic species approach the same value $C_\infty/2$ at the reservoirs. We consider bulk ionic concentrations in the range of $C_\infty \sim 10^{-2}-10^{2}$ mM corresponding to Debye lengths $\lambda \sim 1 - 10^2$ nm, which is non-vanishing in that it is small but not negligible with respect to the confinement size like capillary radii $R \sim 0.1 - 10$ $\mu$m. Here we study the behavior at long time scales such that fluxes of the ions and surface charge relaxes to steady (time-independent) values. More specifically, we are looking at the system after some space-charge relaxation time $\tau > R \lambda/ D_\pm$ \cite{Bazant2004, Janssen2018}, where $D_\pm$ is the diffusivity of the ion species which typically $\sim 10^{-9}-10^{-8}$ m$^2$ s$^{-1}$. In the following subsections we detail the governing equations of our steady state model, the analytical approximations employed to solve the system, and solutions obtained.

\subsection{Governing equations and boundary conditions}  \label{subsecEqns}
We treat the electrolyte as a linear dielectric medium with the $+$ and $-$ ionic concentration fields $c_\pm(r,z)$ at steady state. The electric displacement field is $\varepsilon \mathbf{E}$ and the curl-free electric field $\mathbf{E}=-\nabla\phi$, where $\phi(r,z)$ is the electric potential. The balance of charge is inscribed by Gauss's law as $\nabla \cdot (\varepsilon \mathbf{E}) = e (c_+-c_-)$, where we assume the permittivity of the solution to be isotropic and homogeneous with a constant value. This results in the Poisson's equation for electrostatics,
\begin{equation} \label{eqP} 
      \nabla^2 \phi = -\frac{e}{\varepsilon} (c_+-c_-) \,.
\end{equation}
The ion concentrations considered are dilute such that the solution is modeled as an ideal mixture, and the P\'eclet number Pe $\sim v_0 \lambda/ D_\pm$ is small in our regime of length scales for the characteristic flow speed $v_0$. Thus, the net (volume) fluxes of the ionic species are 
\begin{equation} \label{eqflux1} 
      \mathbf{J}_\pm = - D_\pm \left(\nabla c_\pm  \pm \frac{e}{k_BT} c_\pm \nabla \phi \right) \,,
\end{equation}
where the dominant contributions are provided by the diffusive flux and the electrochemical flux, while the advective contribution, subject to conditions of the aforementioned discussion, is not taken into account. Thus, the conservation of mass $\nabla \cdot \mathbf{J}_\pm =0$ gives the steady state Nernst-Planck equations, 
\begin{equation} \label{eqNP} 
     \nabla^2 c_\pm \pm \frac{e}{k_BT} \nabla \cdot (c_\pm \nabla \phi) = 0 \,.
\end{equation}

We also consider the electrolyte as an incompressible fluid and we operate in the regime of low Reynolds number Re $\sim \rho_0 v_0 \lambda/ \eta$, with $\rho_0$ the mass density and $\eta$ the dynamic viscosity of liquid water. The momentum balance for the fluid with velocity $\mathbf{v}$ is provided by the addition of the electric body force (volume force density) $\mathbf{f}=e (c_+-c_-) \mathbf{E}$ to the Stokes equations. Hence we have the time-independent Stokes equations,
\begin{equation} \label{eqS1} 
    -\nabla P + \eta \nabla^2 \mathbf{v} = e (c_+-c_-) \nabla \phi  \,,
\end{equation}
\begin{equation} \label{eqS2} 
    \nabla \cdot \mathbf{v}=0 \,
\end{equation}
where $P$ is the pressure. 
The coupled system of Eqs.~(\ref{eqP}), (\ref{eqNP}), (\ref{eqS1}) and (\ref{eqS2}) are subject to the following boundary conditions at the capillary wall. First, is the insulating condition for spatially varying surface charge, 
\begin{equation} \label{eqBC1} 
     - \nabla \phi \cdot \hat{n} = \frac{\sigma(z)}{\varepsilon}   \,,  
\end{equation}
where $\hat{n}$ is the unit vector pointing into the liquid and $\sigma$ was defined in \rr{ws}.  
Second, is the spatially varying cation/anion ($\pm$) fluxes coming inwards (from the wall into the fluid domain) and going outwards (from the fluid domain into the wall),
\begin{equation} \label{eqBC2} 
     \mathbf{J}_\pm \cdot \hat{n} = j_\pm (z) \,,
\end{equation}
where the $j_\pm$ were defined in \rr{wjpm}. Additionally, on a solid wall we impose the no-slip and no-penetration boundary condition $\mathbf{v} =\mathbf{0}$. For the purposes of our analytical formulation given below, the capillary is approximated to be of infinte extent in the $z$-direction.

\subsection{Small field perturbations} \label{subsecPeturb}
Consider small non-equilibrium deviations $\delta c_\pm$ of the concentration away from its bulk counterpart $c_\pm=C_\infty/2 + \delta c_\pm$, expected to be valid for small peak amplitudes of the ionic fluxes and surface charge. First we write the dimensionless form of Eqs.~(\ref{eqP}) and (\ref{eqNP}) by employing the dimensionless quantities $\tilde{r} = r/\lambda$, $\tilde{z} = z/\lambda$, $\tilde{R} = R/\lambda$ and $\tilde{k} = \lambda k$, and the dimensionless fields $ \tilde{\phi}= e \phi/k_BT$ and $\tilde{c}_{\pm} = 2 c_\pm/C_\infty$. We introduce fields $\tilde{\rho}=(\tilde{c}_+-\tilde{c}_-)/2$ and $\tilde{s}=(\tilde{c}_++\tilde{c}_-)/2$ and reduce our steady state equations with perturbed fields  $\tilde{\rho} \approx \delta \tilde{\rho}$, $\tilde{s} \approx 1+\delta \tilde{s}$, and $\tilde{\nabla} \tilde{\phi} \approx \tilde{\nabla} \delta \tilde{\phi}$. This gives the leading order equations which we solve analytically with the corresponding boundary conditions (see Appendix~\ref{appenLeq}). We obtain the following expressions for the leading order field perturbations,
\begin{equation} \label{solrho}
    \delta \tilde{\rho} (\tilde{r},\tilde{z}) = {I \big(\sqrt{1+\tilde{k}^2} \ \tilde{r} \big) \left[a_0 \cos(\tilde{k} \tilde{z}) + b_0 \sin(\tilde{k} \tilde{z}))\right]}
\end{equation}
\begin{equation} \label{solphi}
    \delta \tilde{\phi} (\tilde{r},\tilde{z}) = {I (\tilde{k} \tilde{r} ) \left[a_1 \cos(\tilde{k} \tilde{z}) + b_1 \sin(\tilde{k} \tilde{z}))\right]}   -  \delta \tilde{\rho} \,,
\end{equation}
where
\be \label{I}
I(  K \tilde{r} ) = \frac{I_0 ( K \tilde{r} )}{K \ I_1 (K \tilde{R} )}, 
\ee
$I_{0}$ and $I_1$ are the modified Bessel functions of first kind,  
\begin{equation} \label{solCoefa0}
    a_0 = \delta_+ \cos{\alpha_+} -\delta_- \cos{\alpha_-} - \delta_\sigma \cos{\beta} \,,
\end{equation}
\begin{equation} \label{solCoefb0}
   b_0= -\delta_+ \sin{\alpha_+} + \delta_- \sin{\alpha_-} + \delta_\sigma \sin{\beta} \,,
\end{equation}
\begin{equation} \label{solCoefa1}
    a_1 = \delta_+  \cos{\alpha_+} -\delta_- \cos{\alpha_-} \,, 
\end{equation}
\begin{equation} \label{solCoefb1}
  b_1= - \delta_+  \sin{\alpha_+} + \delta_- \sin{\alpha_-} \,,
\end{equation}
with the relative (dimensionless) amplitudes of flux and charge
\begin{equation} \label{deltas}
    \delta_\pm= \frac{j_{0\pm} \lambda}{D_\pm C_\infty} \,, \ \ \ \  \delta_\sigma= \frac{\sigma_0 e \lambda}{\varepsilon k_B T}.
\end{equation}
taken to be small (Appendix~\ref{appenLeq}). 

Eq. \rr{solphi} already clarifies the origin of the effect. The commensurate electric field depends \emph{only} on the surface fluxes $j_{0\pm}$ (the last term in \rr{solphi} does not affect the flow as it is absorbed into the pressure). Without surface flux activity the effect is absent. Below, employing the corresponding body force, we quantify how the flow can acquire circulatory or unidirectional character. 

\subsection{Self-induced body force}  \label{subsecForce}
The body force $\mathbf{f}= (f_r, 0 ,f_z)$ is self-generated as a consequence of the flux-charge boundary mechanism, where the radial and axial components in dimensionless form are
\begin{equation}
 \tilde{f}_r = -  \tilde{\rho} \ \partial_{\tilde{r}} \tilde{\phi} \,, \ \ \ \ 
 \tilde{f}_z = - \tilde{\rho} \ \partial_{\tilde{z}} \tilde{\phi} \,,
\end{equation}
where $\tilde{f}_{r,z} = \lambda f_{r,z} /k_BT C_\infty$. Retaining only the part of the force that cannot be absorbed into the pressure in Eq.~(\ref{eqS1}), the axial component becomes
\be\label{eqForce}
\begin{split} 
 \tilde{f}_z \approx \frac{\tilde{k}}{2}I \big(\sqrt{1+\tilde{k}^2} \tilde{r} \big) \ I(\tilde{k} \tilde{r} ) \bigg[h_0 &+ h_c \cos(2 \tilde{k} \tilde{z})  \\&+ h_s \sin(2 \tilde{k} \tilde{z}))\bigg] \,,
\end{split}    
\ee
where $I(K \tilde{r} )$ was defined in \rr{I} and  
\begin{widetext}
\begin{equation} \label{eqh0}
    h_0 = \delta_\sigma \delta_+ \sin(\beta-\alpha_+) -\delta_\sigma \delta_- \sin(\beta-\alpha_-) \,,
\end{equation}
\begin{equation} \label{eqhC}
    h_c  = \delta_+^2 \sin(2\alpha_+) + \delta_-^2 \sin(2\alpha_-) - 2 \delta_+ \delta_- \sin(\alpha_+ +\alpha_-)  - \delta_\sigma \delta_+ \sin(\beta+\alpha_+) +\delta_\sigma \delta_- \sin(\beta+\alpha_-) \,,
\end{equation}
\begin{equation} \label{eqhS}
    h_s = \delta_+^2 \cos(2\alpha_+) + \delta_-^2 \cos(2\alpha_-) - 2 \delta_+ \delta_- \cos(\alpha_+ +\alpha_-) - \delta_\sigma \delta_+ \cos(\beta+\alpha_+) +\delta_\sigma \delta_- \cos(\beta+\alpha_-) \,.
\end{equation}
\end{widetext}
The body force expression (\ref{eqForce}) clarifies the character of the resultant flow. The first term in the square brackets of Eq.~(\ref{eqForce}) leads to a force always pointing in the same direction and gives rise to a zero mode (unidirectional) velocity, as it will be discussed below. The second and third terms in the square brackets of Eq.~(\ref{eqForce}) lead to the formation of vortices at every half-period of the wall flux and charge variations. The symmetry properties of the resulting velocity field discussed in Sec.~\ref{secSym} are also evident in Eq.~(\ref{eqForce}), that is, under the spatial and domain inversion this body force changes sign. Thus, it gives rise to velocity fields that will also change sign as predicted in Eq.~\rr{sym}.

\subsection{Long wave mode}  \label{subsecLong}
To obtain some analytical insight into the effect we continue by performing the long-wavelength approximation to the governing fields. The velocity field $\mathbf{v}=(v_r,0,v_z)$ can be expressed with respect to the (dimensionless) Stokes streamfunction $\psi(r,z)$ in the form
\begin{equation} \label{eqVcomp}
    \tilde{v}_r=\frac{1}{\tilde{r}}\partial_{\tilde{z}} \tilde{\psi} \,, \ \ \ \  \tilde{v}_z= - \frac{1}{\tilde{r}}\partial_{\tilde{r}} \tilde{\psi} \,,
\end{equation}
satisfying the isochoric constraint in Eq.~(\ref{eqS2}), and with $\tilde{v}_{r,z}= \eta v_{r,z}/ k_BT C_\infty \lambda$. Taking the curl on both sides in the Stokes equations (Eq.~(\ref{eqS1})) leads to the vorticity equation \cite{Happel1983, Landau1987},
\begin{equation} \label{eqVort}
    \tilde{\mathcal{D}}^4 \tilde{\psi} = \tilde{r} ( \partial_{\tilde{z}} \tilde{\rho} \ \partial_{\tilde{r}} \tilde{\phi} -\partial_{\tilde{r}} \tilde{\rho} \ \partial_{\tilde{z}} \tilde{\phi} ) \,.
\end{equation}
where the operator $\tilde{\mathcal{D}}^4 \tilde{\psi}= \tilde{\mathcal{D}}^2 (\tilde{\mathcal{D}}^2 \tilde{\psi})$ and
\begin{equation}
    \tilde{\mathcal{D}}^2  \tilde{\psi} =\left(\partial_{\tilde{r}}^2-\frac{1}{\tilde{r}}  \partial_{\tilde{r}} + \partial_{\tilde{z}}^2 \right)  \tilde{\psi} \,.
\end{equation}
The streamfunction satisfies the boundary conditions (no-slip and no-penetration velocity field)
\begin{equation} \label{eqBCns}
     \tilde{\psi} \ |_{\tilde{r}=\tilde{R}} = 0 \,, \ \ \ \   \partial_{\tilde{r}} \tilde{\psi} \ |_{\tilde{r}=\tilde{R}} =0\,.
\end{equation}

Here we look at the ionic flux and surface charge variation patterns in Eqs~(\ref{wjpm}) and (\ref{ws}), respectively, with a large (but finite) wavelength compared to the Debye length, like $l \sim 0.1 - 10^2$ $\mu$m for instance. Keeping terms up to order $\mathcal{O}(\tilde{k}^{-1})$, we get the analytical expressions for the leading order streamfunction (see Appendix~\ref{appenLeq})
\begin{equation} \label{eqStream}
     \tilde{\psi} (\tilde{r}, \tilde{z}) \approx \Psi_0(\tilde{r}) h_0 + \Psi_1(\tilde{r}) \left[h_c \cos(2\tilde{k}\tilde{z}) + h_s \sin(2\tilde{k}\tilde{z})\right] \,,
\end{equation}
where $ \Psi_0(\tilde{r})$ and $ \Psi_1(\tilde{r})$ are radially-dependent functions whose explicit form is displayed in Appendix~\ref{appenCoeff}. It is clear that the streamfunction \rr{eqStream} displays two distinct behaviors. The first term on the right hand-side gives rise to a unidirectional velocity component (a zero mode), which is a natural consequence of the respective body force (first term in the square brackets of Eq. \rr{eqForce}), cf. panel (c) of Fig. \ref{fig2}. The remaining two terms in the streamfunction \rr{eqStream} give rise to circulatory behavior, in accordance with the last two terms in the square brackets of Eq. \rr{eqForce}, cf. panel (b) of Fig. \ref{fig2}. Cross term effects have appeared before in the literature \cite{Ristenpart2008}, for a two-dimensional channel, albeit with respect to a formulation different than ours. The electric body force in their case was linearized with respect to a basic state (of uniform surface charge and ionic flux conditions) and thus gave rise to circulating flows only without any zero mode. In contrast, here we establish the existence of a zero mode velocity component \emph{and} circulatory flows, the latter as a consequence of wall flux activity alone. The resulting flow field and net velocity from our stream function are discussed in the next sections, where we analyze the selection rules revealing their specific dependencies on key parameters.

\section{Flux-charge phase difference} \label{secFlow}
Active-charged patterns on the capillary wall produce a variety of flow states with a rich dependence on the flux and charge phase differences resulting in various selection rules for $\alpha_\pm$ and $\beta$. Using the stream function Eq.~(\ref{eqStream}) derived above, we find the leading order radial and axial fluid velocity components from Eq.~(\ref{eqVcomp}) as
\begin{equation} \label{eqVr}
     \tilde{v}_r (\tilde{r}, \tilde{z}) \approx  \frac{2 \tilde{k}}{\tilde{r}} \Psi_1(\tilde{r}) [ h_s \cos(2\tilde{k}\tilde{z}) -  h_c \sin(2\tilde{k}\tilde{z})] \,,
\end{equation}
\begin{equation}  \label{eqVz}
    \tilde{v}_z (\tilde{r}, \tilde{z}) \approx U(\tilde{r}) \ h_0 + W(\tilde{r}) [h_c \cos(2\tilde{k}\tilde{z}) +  h_s \sin(2\tilde{k}\tilde{z})] \,,
\end{equation}
where we denote
\begin{equation}
    U(\tilde{r})= - \frac{1}{\tilde{r}} \partial_{\tilde{r}} \Psi_0 \,, \ \ \ \  W(\tilde{r})= - \frac{1}{\tilde{r}} \partial_{\tilde{r}} \Psi_1 \,
\end{equation}
and $ \Psi_0(\tilde{r})$ and $ \Psi_1(\tilde{r})$ were defined in Appendix~\ref{appenCoeff}. The factors $U(\tilde{r}) $ and $W(\tilde{r}) $ are odd with respect to $\tilde{k}$. Thus, under spatial inversion the velocity $ \tilde{v}_z$ changes sign. The form of Eq. \rr{eqVz} was predicted in Eq. \rr{sym} by symmetry principles. 
  
The $z$-component of the velocity field in Eq. \rr{eqVz} is composed of linear combinations of plug-like (unidirectional) profiles $U(\tilde{r}) $ and oscillating (circulatory) profiles $W(\tilde{r}) $. The shapes of these unidirectional and circulatory profiles are determined by $\Psi_{0,1}$ given in Eq.~(\ref{eqStream0}) and (\ref{eqStream1}) as depicted in Fig.~\ref{fig2}(a), and their relative contributions rely on the coefficient factors $h_{0,c,s}$ which are given by Eqs.~(\ref{eqh0})-(\ref{eqhS}). If the amplitudes of both fluxes are set to zero $j_{0 \pm}=0$ (no activity) then all these coefficient factors vanish, resulting in no flow. Thus at least one of the fluxes needs to be active for inducing non-vanishing flows. We now proceed to discuss specific scenarios that lead to these flows with the cationic flux centered at the origin, that is we set $\alpha_+=0$ from here on out. 

A basic condition that leads to a steady flow state, even with vanishing surface charge $\sigma_0=0$ ($\delta_\sigma=0$), is when only one of the ionic flux, say the cationic flux is active $j_{0+} \neq 0$ ($\delta_+ \neq 0$) while the anionic flux is set to zero $j_{0-} = 0$ ($\delta_- = 0$). Under this condition, only the circulatory part remains as $h_s \neq 0$, but the unidirectional part vanishes since $h_0=0$. In the converse situation with only the anionic flux active and centered at origin instead, leads to the exact same circulatory states. Another case with no surface charge that also results in such circulatory flows is when both fluxes are active $j_{0 \pm} \neq 0$ given that their relative amplitudes are not the same which means $\delta_+ \neq \delta_-$. But if $\delta_+ = \delta_-$ then the cationic-anionic flux phase difference needs to be $\alpha_- \neq n \pi/2$ (for some integer $n$) such that it gives non-vanishing circulating flows. However, in all these scenarios with vanishing surface charge, only circulatory states can arise. We will now look what happens when a flux-charge phase difference is introduced with non-vanishing surface charge pattern $\sigma_0 \neq 0$ ($\delta_\sigma \neq 0$).

\subsection{Case with only cationic activity}
Lets first look at the effect of just the phase difference $\beta$ between cationic flux and surface charge by setting the anionic flux to zero $j_{0-}=0$, that is $\delta_-=0$ in Eqs.~(\ref{eqh0})-(\ref{eqhS}). Here if we also set flux-charge amplitudes such that $\delta_+ = \delta_\sigma$, the coefficients of the flow velocity turn out as $h_0 \sim \sin(\beta)$, $h_c \sim - \sin(\beta)$, and $h_s \sim 1  - \cos(\beta)$. Then when the flux and charge patterns are in-phase (same phase) $\beta=2 n \pi$, all the coefficients vanish leading no flow as $v_{r,z} \approx 0$ implying that the fluid becomes stagnant. But if they are out-of-phase (opposite phase) $\beta = (2 n+1) \pi$, the unidirectional component vanishes and the axial velocity becomes $\tilde{v}_z \sim W \sin(\tilde{k}\tilde{z})$. Hence in this case, we only develop fully circulatory flow states of toroidal vortex patterns that alternate in direction with profile shapes of $\pm W$ across the capillary (see Fig.~\ref{fig2}(b)). However, if the flux-charge patterns are offset by an arbitrary phase difference that is not of the same or opposite phase, then a unidirectional contribution emerges. For instance, when $\beta=\pi/2$ the unidirectional contribution is maximal as $\tilde{v}_z \sim U + W [\sin(\tilde{k}\tilde{z})-\cos(\tilde{k}\tilde{z})]$ leading to a unidirectional dominant flow diminishing the vortex patterns (see Fig.~\ref{fig2}(c)). But if the relative contributions of unidirectional and circulatory components are comparable, like when $\beta=3\pi/4$, it results in intermediate states with alternating expanded and contracted vortex patterns where along the capillary it has alternating profile shapes $U \pm W$ (see Fig.~\ref{fig2}(d)).   

\begin{figure*}[t!]
    \centering
    \includegraphics[width=1.0\textwidth]{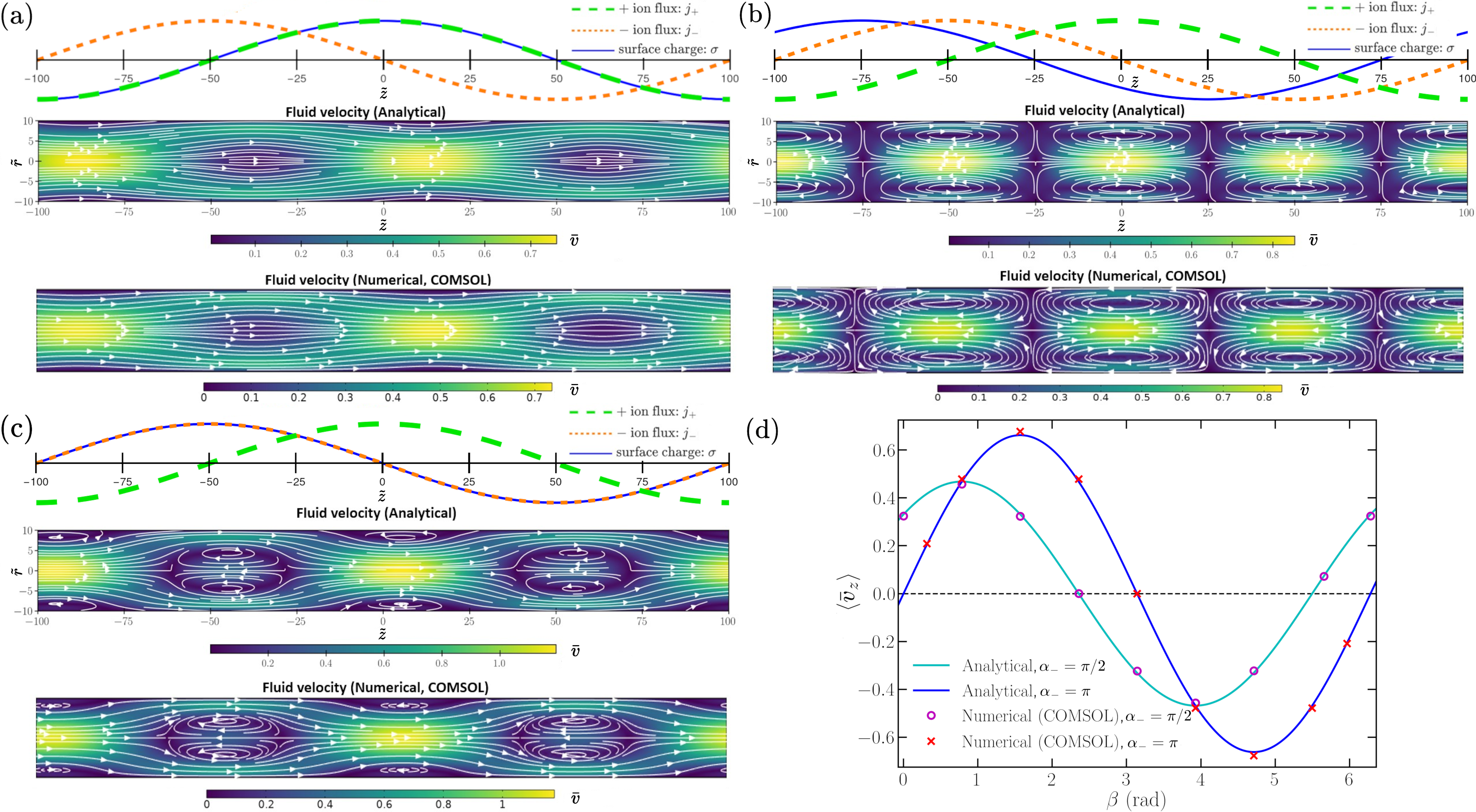}
    \caption{Self-generated electrokinetic flows from flux-charge patterns with both cationic and anionic activity. The stream plots of fluid velocity in $(\tilde{r},\tilde{z})$ space obtained analytically from Eq.~(\ref{eqVr}) and~(\ref{eqVz}) and numerically using {\it COMSOL} (Appendix~\ref{appenNum}), show good agreement. The modulation pattern configurations employ $\alpha_+=0$ and $\alpha_-=\pi/2$ at different flux-charge phase difference, leading to (a) unidirectional dominant flow at $\beta=0$, (b) fully circulatory flow at $\beta=3 \pi/4$ and (c) intermediate flow state at $\beta=\pi/2$. The corresponding modulation forms of the cationic (+) flux, the anionic (-) flux and the surface charge are shown above. The arrows of streamlines indicate the flow direction, and the color map indicates the velocity magnitude where $\tilde{r}=r/\lambda$, $\tilde{z}=z/\lambda$ and  $\bar{v}= v/\lambda$ s$^{-1}$. (d) Averaged capillary flow velocity plot with the wavelength averaged z-component velocity as a function of flux-charge phase difference $\beta$ where $\langle \bar{v}_z \rangle=\langle v_z \rangle/\lambda$ s$^{-1}$ with $\alpha_+=0$ and $\alpha_-$ as indicated, comparing the analytical result given by Eq.~(\ref{eqAvgVz}) with the numerical result computed in {\it COMSOL} (Appendix~\ref{appenNum}). The quantities are given in terms of the Debye length $\lambda$ for bulk concentration $C_\infty=1$ mM. We use here capillary radius $R=10 \lambda$ and wavelength $l=200 \lambda$ with flux-charge peak amplitudes such that $\delta_{\pm} \approx \delta_\sigma \approx 3.2 \times 10^{-4}$ for ions with same diffusivity $D_\pm = 10^{-9}$ m$^2$s$^{-1}$.}
    \label{fig3}
\end{figure*}

This effect arising from the flux-charge phase difference can be physically understood as spatial mismatch or defects appearing in the surface structure that constantly injects electrochemical gradients and triggers fluid flow as a response. For instance, here with the anionic flux set to zero, the parameter $\beta$ represents the degree of mismatch between the sites of peak cation in-flux (out-flux) and peak positive (negative) surface charge. When in-phase like $\beta=0$, the peak sites are in perfect alignment and the patterns fully overlap such that the dislocated charges balances out. As a result, body force inducing electrochemical gradients vanish as fluid settles to rest. In the case of an anti-alignment of the peak sites like for $\beta=\pi$, that is cation in-flux peak matches with negative surface charge peak and vice versa. This type of opposing overlap incurs a constant redistribution of charge and ends up in a steady circulating flow. However an offset with an arbitrary misalignment of the flux-charge peaks like $\beta=\pi/2$, breaks spatial symmetry of translation-reflection and leads to transverse body forces that persist, invoking steady flows with a preferred net direction. Moreover, in a biological context this can be envisioned as a scenario of elongated cell membranes with an alternating sequence of gated ion pumps that facilitate in-and out-flux of cations, and a lipid bilayer surface that carries heterogeneous distribution of charge. This can also be realized in a laboratory construction by periodic coatings of ion-absorbing and ion-releasing chemical reactants grafted on micro-channels or capillary walls in addition with alternating induction of surface charges. 

Additionally we note that in the converse case of only anionic activity ($j_{0-} \neq 0$) with cationic flux now set to zero ($j_{0+}=0$), the flow directions are reversed and states are changed according to $\beta$ with the peak anionic flux placed at the center. Specifically, in Eqs.~(\ref{eqh0})-(\ref{eqhS}) for $\delta_+=0$ and $\alpha_-=0$, we obtain the two coefficients with flipped signs $h_{0,c} \rightarrow - h_{0,c}$, whereas the other one goes as $h_s \rightarrow 2-h_s$. This means that when the ionic activity is swapped (from cationic only to anionic only), for a flux-charge phase difference like $\beta=\pi/2$ of the former case (Fig.~\ref{fig2}(c)), the unidirectional and the circulatory parts in the latter case turn to the opposite directions reversing the flow. However, for the intermediate state seen previously at $\beta=3\pi/4$ (Fig.~\ref{fig2}(c)), the corresponding reversed scenario with similar vortex structure would now be that of $\beta=\pi/4$ and not $\beta=3 \pi/4$ here. The swapping of ionic activity can also leads to a shift in flow states. Furthermore, the consequence of being in-phase or out-of-phase are switched here, namely for same phase $\beta=2 n \pi$ there is only circulatory flows and instead for opposite phase $\beta=(2 n +1)\pi$ there are no flows.

\subsection{Case with both cationic and anionic activity}
Next we look at the scenario where both fluxes are non-zero and examine the resulting flow patterns induced by the phase differences. Figures~\ref{fig3}(a)-(c) illustrates the fluid velocity when ion diffusivity and flux amplitudes are identical, $D_+=D_-$ and $j_{0+}=j_{0-}$ such that the relative flux and charge amplitudes are all the same $\delta_+=\delta_-=\delta_\sigma$. In this case if the cationic and anionic fluxes are in-phase $\alpha_-=2 \pi n$, then there is no unidirectional component as $h_0 = 0$ in Eq.~(\ref{eqh0}) regardless of what the flux-charge phase difference $\beta$ is. Thus leading to only circulatory flows producing the toroidal vortex patterns. However, if the $\pm$ fluxes are offset with some arbitrary $\alpha_-$, we get a non-vanishing unidirectional component depending on $\beta$ and results in a variety of composite flow states. For instance when $\alpha_-= \pi/2$, the unidirectional part dominates at $\beta=0$ diluting the circulatory vortex patterns (Fig.~\ref{fig3}(a)), whereas in contrast, unidirectionality subsides at $\beta=3 \pi/4$ giving fully circulatory flows with amplified vortex patterns (Fig.~\ref{fig3}(b)). We observe the alternating vortex patterns at intermediary flow states here for $\beta=\pi/2$ (Fig.~\ref{fig3}(c)). In addition, we validate our analytical results with numerical finite element simulations as shown in Figs.~\ref{fig3}(a)-(c) using the {\it COMSOL Multiphysics} software (see Appendix~\ref{appenNum} for details).

With both ionic fluxes being non-zero, the phase difference $\alpha_-$ can be thought of as an extent of spatial separation between two types of active sites, one that produces cationic flux and the other produces anionic flux. For the special case when the peak $\pm$ fluxes are fully aligned, like at $\alpha_-=0$, this can be taken as scenario of a surface activity that simultaneously produces fluxes of both ionic species at the same site. If the relative cationic-anionic flux amplitudes are also the same, then this can only result in circulating flows independently of how the surface charge pattern is offset. Here the effect of the cationic flux-charge phase difference $\beta$ discussed previously (in Sec.~\ref{secFlow}A) that gives rise to the unidirectional flow gets counteracted by the exact opposite effect from anionic flux-charge phase difference which in this case is also $\beta$. Thus the effect of flux-charge phase difference is suppressed when the cites of cationic-anionic activity (in- and out-flux) fully coincide. However for some arbitrary spatial discrepancies in the peak cationic-anionic active sites such as with $\alpha_-=\pi/2$ that is not in-phase, the opposing effects are now mismatched with some non-vanishing net unidirectional component. The degree of this mismatch can then be tuned by the offset of the charge pattern, resulting again in a spectrum of flow states. 

Taken together, a spectrum of steady states from circulatory, unidirectional and intermediate flows are obtained through active-charged modulation patterns on confinement boundaries. Flow states with a non-vanishing unidirectional component emerge when ionic flux and surface charge patterns are offset by a flux-charge phase difference attributed to a spatial mismatch in the peak sites of flux and charge distributions. This can be achieved with just one ionic (cationic or anionic) activity (Sec.~\ref{secFlow}A) as well as for the case with both (Sec.~\ref{secFlow}B). In the former case, the full spectrum of flow states are produce when the flux and charge patterns mismatched by a phase difference such that they are neither of the same or opposite phase $\beta \neq n \pi$. Moreover, swapping out the type of ionic activity here (from cationic to anionic, or vice versa) leads to a complete flow reversal or a change of the flow patterns for a fixed phase difference. In the latter case, the simultaneous presences of equal cationic and anionic flux at the same cite suppresses unidirectional flow, and an offset between the two fluxes $\alpha_- \neq 2 n \pi$ is required to recover the full spectrum of flow states. We focused in this section on the effect of phase difference parameters by setting the relative flux and charge amplitudes as equivalent. We will explore in the Sec.~\ref{secEffect} the effect of having different relative flux magnitudes by the means of ion diffusivity mismatch. In the following section we characterize the averaged fluid flow across the capillary and provide estimates for the net unidirectional velocity obtained as a function of the key parameters.

\begin{table*}[t!]
\begin{tabular}{ |c|c|c|c|c| } 
\hline
\multirow{3}{8em}{Debye length \\ (bulk ion \\ concentration)} & \multirow{3}{8em}{Flux amplitude \\ $j_{0+}$ (mM m s$^{-1}$) } & \multirow{3}{8em}{Charge amplitude  
 \\ $\sigma_0$ (C m$^{-2}$) }  & \multirow{3}{8em}{Relative amplitudes \\ (dimensionless) \\ for $\delta_+ \approx \delta_\sigma$} & \multirow{3}{12em}{Average (peak) flow velocity \\ for radius $R = 0.5$ $\mu$m and \\   wavelength $l = 1$ $\mu$m  }  \\
& & & &  \\
& & & &  \\
\hline
\multirow{4}{8em}{$\lambda \approx 14$ nm \\ ($C_\infty=1$ mM)}  & \multirow{2}{8em}{$2.3 \times 10^{-4}$} & \multirow{2}{8em}{$4.2 \times 10^{-6}$} & \multirow{2}{8em}{$\delta_{+,\sigma} \approx 3.2 \times 10^{-3}$}  & \multirow{2}{12em}{$ \langle v_z \rangle \approx 0.1$ $\mu$m s$^{-1}$}  \\ 
& & & & \\
& \multirow{2}{8em}{$7.3 \times 10^{-3}$} & \multirow{2}{8em}{$1.3 \times 10^{-4}$} & \multirow{2}{8em}{$\delta_{+,\sigma} \approx 0.1$}  & \multirow{2}{12em}{$\langle v_z \rangle \approx 10^{3}$ $\mu$m s$^{-1}$} \\ 
& & & & \\
\hline
\multirow{4}{8em}{$\lambda \approx 1.4$ nm \\ ($C_\infty=10^2$ mM)}  & \multirow{2}{8em}{$2.3 \times 10^{-1}$} & \multirow{2}{8em}{$4.2 \times 10^{-5}$} & \multirow{2}{8em}{$\delta_{+,\sigma} \approx 3.2 \times 10^{-3}$}  & \multirow{2}{12em}{$ \langle v_z \rangle \approx 1$ $\mu$m s$^{-1}$}  \\ 
& & & & \\
& \multirow{2}{8em}{$7.3$} & \multirow{2}{8em}{$1.3 \times 10^{-3}$} & \multirow{2}{8em}{$\delta_{+,\sigma} \approx 0.1$}  & \multirow{2}{12em}{$\langle v_z \rangle \approx 10^4$ $\mu$m s$^{-1}$} \\ 
& & & & \\
\hline
\end{tabular}
\caption{\label{table1} Estimates of the average unidirectional (peak) velocity given by Eq.~(\ref{eqAvgVz}), with only cationic flux active for radius $R = 0.5$ $\mu$m and wavelength of $l = 1$ $\mu$m, evaluated at different Debye lengths $\lambda$ (for corresponding bulk ion concentrations $C_\infty$) and peak amplitudes of cationic flux ($j_{0+}$ in units of m s$^{-1}$) and surface charge ($\sigma_0$ in units of m$^{-2}$). The values of the flux and charge amplitudes taken here are small, that is the (dimensionless) amplitudes in Eq.~(\ref{deltas}) are $\delta_{+} \approx \delta_\sigma  \lesssim 0.1$ and $kR \approx 3$.}
\end{table*}

\section{Average unidirectional velocity} \label{secAvg}
We quantify the net unidirectional flow inside the capillary with the averaged ($z$-component) \emph{dimensional} velocity 
$  \langle v_z \rangle$
taken over a wavelength sized cylindrical volume defined as
\begin{equation} \label{eqAvDef}
    \langle v_z \rangle = \frac{1}{\pi R^2 l} \int_{-l/2}^{l/2} \int_{0}^{R} 2 \pi r v_z \ dr \ dz \,.
\end{equation}
The leading order term in the long wavelength approximation, it can be written explicitly in terms of key parameters in the form
\begin{equation} \label{eqAvgVz}
\begin{split}
    \langle v_z \rangle \approx  \frac{\sigma_0 e R^2 l}{2 \pi \varepsilon \eta} \left( \frac{\lambda}{R} \right)^3 \mathscr{L}_{\textrm{cyl}}(R/\lambda)  \bigg[ & \frac{j_{0+}}{D_+} \sin(\beta - \alpha_+) \\ & - \frac{j_{0-}}{D_-} \sin(\beta-\alpha_-) \bigg] \,,
\end{split}    
\end{equation}
where we defined the ``cylindrical'' Langevin function 
\be\label{eqAf}
\mathscr{L}_{\textrm{cyl}} (x) =  \frac{I_0(x)}{I_1(x)}  - \frac{2}{x}
\ee
whose asymptotics are $\mathscr{L}_{\textrm{cyl}} (x)= \frac{x}{4} +O(x^3) $ as $x\rightarrow 0$ and $\mathscr{L}_{\textrm{cyl}} (x)= 1 -\frac{3}{2x}+O(x^{-2}) $ as $x\rightarrow \infty$. The averaging procedure carried-out in Eq. \rr{eqAvgVz}, eliminated the circulatory flow contributions and only retained the zero mode. The magnitude is quadratic with respect to flux and charge amplitudes of the form $j_{0 +}\sigma_0$ and $j_{0 -}\sigma_0$, and scales with the wavelength $l$. This net flow vanishes in (infinitesimally) thin Debye layer limit, namely as the Debye length to capillary radius ratio $\lambda/R \rightarrow 0$ its magnitude decreases rapidly to zero as the factor $ \left( \frac{\lambda}{R} \right)^3 \mathscr{L}_{\textrm{cyl}}(R/\lambda) \rightarrow 0$. However, at a Debye length that is sizable enough we can still get a significantly large magnitude for the averaged speed.  

We find that the averaged velocity varies sinusoidally as function of the phase differences which is shown with the corresponding numerically obtained values in Fig.~\ref{fig3}(d), for the case of equal ionic flux to diffusivity ratio $j_{0+}/D_+ = j_{0-}/D_-$. The peak magnitude and direction reversing of the net velocity here are different depending on phase difference combinations. For instance with $\alpha_+=0$, in the case when $\alpha_-=\pi/2$ the velocity goes as $\sim \sqrt{2} \sin(\beta+\pi/4)$ and in contrast for $\alpha_-=\pi$ it goes as $\sim 2 \sin \beta$, where their magnitudes differ by a factor of $\sqrt{2}$ and direction switching happens at a distinct $\beta$ values (Fig.~\ref{fig3}(d)). This means that in this case even when the relative cationic-anionic flux magnitudes are equal, there is an optimal pattern configuration that gives the maximal magnitude of unidirectional velocity. In our discussion earlier in Sec.~\ref{secFlow}B, we found that the cationic and anionic fluxes are need to be offset by some $\alpha_- \neq 2 n \pi$ to give rise to non-vanishing unidirectional flow, and here we add that when they are particularly in the opposite phase $\alpha_-=(2n+1) \pi$ its magnitude is maximized. This configuration can be thought of as a situation where the site of cationic in-flux and anionic out-flux coincide, or in another sense, there are biased active sites that allows for ion-selective fluxes. In this optimal case the effect coming from the cationic flux-charge phase difference is exactly superimposed with the aligning effect of the anionic counterpart, and thus the unidirectional flow is amplified.  

We now estimate the magnitude of this net flow velocity based on our analytic approximation for the case of only cationic flux active ($j_{0-}=0$) in Eq.~(\ref{eqAvgVz}), by expressing its peak value ($\beta-\alpha_+=\pi/2$) as
\be \label{eqVdel}
\langle v_z \rangle \approx   \delta_+ \delta_\sigma \ V 
\ee
where
\be \label{eqVmax}
V = \frac{k_BT C_\infty \lambda} {\eta} \ \frac{l}{2 \pi R} \ \mathscr{L}_{\textrm{cyl}}(R/\lambda) 
\ee
In particular, we use the values for the amplitudes of cationic flux and surface charge such that they are relatively small in the long-wavelength limit, meaning the relative (dimensionless) amplitudes $\delta_{+,\sigma}<1$ and $kR>2 \delta_+$ (see Appendix~\ref{appenApprox}). Under these criteria, when the Debye length is smaller (or bulk concentration is higher), larger values of flux and charge amplitudes can be adopted yielding larger speeds. For example, at radius $R = 0.5$ $\mu$m and wavelength $l = 1$ $\mu$m, when $\lambda \approx 14$ nm we have $\mathscr{L}_{\textrm{cyl}}(R/\lambda) \approx 0.85$. We get the limiting speed $V \approx 0.1$ m s$^{-1}$ leading to the average unidirectional speeds like $\langle v_z \rangle \approx 1$  mm s$^{-1}$. At even smaller Debye lengths, if we keep the same (absolute) amplitudes of flux and charge the speed drops by three orders of magnitude. However, if instead use amplitudes that are relatively small now with respect to the smaller Debye length then we get larger speeds, and with much stronger flux. Table~\ref{table1} shows the estimates for this speed at various ranges of parameter values. Thus, given such Debye lengths and amplitudes of ionic flux and surface charge, we can get here average (peak) flow speeds ranging from orders of microns up to millimeters per second.

\begin{figure*}[t!]
	\centering
	\includegraphics[width=0.95\textwidth]{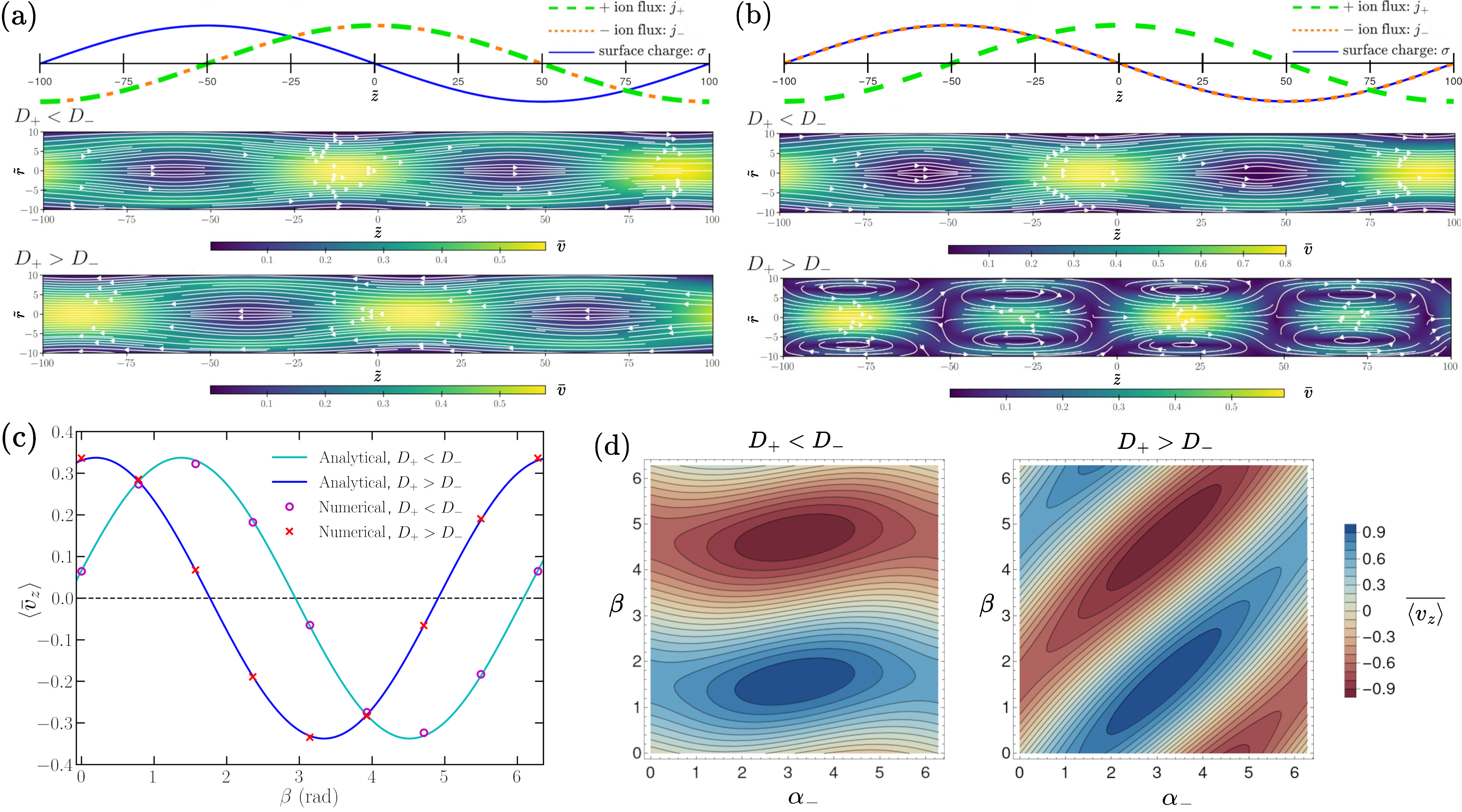}
	\caption{Effect of ionic diffusivity mismatch. The stream plots of steady state fluid velocity in $(\tilde{r},\tilde{z})$ obtained analytically from Eq.~(\ref{eqVr}) and~(\ref{eqVz}) with different ion diffusivity $D_+ \neq D_-$ mismatch with $\alpha_+=0$ for (a) $(\alpha_-,\beta)=(0,\pi/2)$ and (b) $(\alpha_-,\beta)=(\pi/2,\pi/2)$. The corresponding modulation forms of the cationic (+) flux, the anionic (-) flux and the surface charge are shown above. The arrows of streamlines indicate the flow direction and the color map indicates the velocity magnitude given in dimensionless units as in Fig.~\ref{fig2}. (c) Plot of the wavelength averaged z-component velocity as a function of flux-charge phase difference $\beta$ at $(\alpha_+,\alpha_-)=(0,\pi/2)$ where $\langle \bar{v}_z \rangle=\langle v_z \rangle/\lambda$ s$^{-1}$ as in Fig.~\ref{fig3}, comparing the analytical approximation given by Eq.~(\ref{eqAvgVz}) with the numerical computed values (using  {\it COMSOL}, Appendix~\ref{appenNum}) and leading to an good agreement. (d) The landscape (contour plot) of normalized average velocity $\bar{\langle v_z \rangle}=\langle v_z \rangle/|\langle v_z \rangle|_{\text{max}}$ as function of $(\alpha_-,\beta)$ at $\alpha_+=0$ for different diffusivity mismatch. We use here $R=10 \lambda$ and $l=200 \lambda$ for ions with same flux amplitudes $j_{0+} = j_{0-}$, and we use $D_- = 5 D_+$ for $D_+ < D_-$ and $D_+ = 5 D_-$ for $D_+ > D_-$.}
	\label{fig4}
\end{figure*}

The fast directed flows arise at long wavelengths due to strong lateral body forces that are produced when high ionic fluxes constantly disrupt a tightly bound double layer. This can be seen from the averaged body force component of Eq.~(\ref{eqForce}) that drives the unidirectional flow, which in the long wavelength mode can be written as
\begin{equation}
\begin{split}
    \langle f_z \rangle \approx \frac{\sigma_0 e l}{\pi \varepsilon} \left(\frac{\lambda}{R} \right)^2 \bigg[ & \frac{j_{0+}}{D_+} \sin(\beta - \alpha_+) \\ & - \frac{j_{0-}}{D_-} \sin(\beta-\alpha_-) \bigg] \,,
\end{split}    
\end{equation}
where the averaging is defined as in Eq.~(\ref{eqAvDef}). Assuming $j_{0-}=0$, the averaged driving body force scales as 
\begin{equation}
    \langle f_z \rangle \sim \frac{j_{0+}\sigma_0 e l}{\pi \varepsilon D_+} \left(\frac{\lambda}{R} \right)^2.
\end{equation}
Thus, in a manner analogous to the averaged flow velocity, for a non-vanishing Debye length the magnitude of the driving force is quadratic with respect to the amplitudes of flux and charge. This prominent effect takes hold as a consequence of significant ionic elevation and depletion in the double layer that keeps knocking it off the equilibrium configuration. Self-induce ionic body forces then persist throughout long ranges maintaining fast transverse fluid motion.

\section{Ion diffusivity mismatch} \label{secEffect}

We examine in this section how changes in key physical parameters affects flow field and net velocity. We focus here on the effect of having mismatch in diffusivity of the ionic species, $D_+ \neq D_-$ and compare the two cases $D_+<D_-$ and $D_+>D_-$. First we note that, in contrast to the case discussed in Sec.~\ref{secFlow}B, when a difference in ionic diffusivity (or more generally a difference in ionic flux-diffusivity ratio $j_{0+}/D_+ \neq j_{0-}/D_-$) is introduced even when the cationic and anionic fluxes are in-phase, like at $\alpha_-=0$, a unidirectional flow emerges when offset by the flux-charge phase difference. 

Along with the magnitude of the net flow velocity, its direction also depends on the difference in the diffusivity ratio $(j_{0+}D_--j_{0-}D_+)/D_+D_-$ as can be seen from Eq.~(\ref{eqAvgVz}) for fluxes of same phase $\alpha_-=2 n \pi$. If we switch here from the $D_+<D_-$ case to the $D_+>D_-$ case, or vice versa, the unidirectional component must flip sign and hence the net flow reverses in direction like shown in Fig.~\ref{fig4}(a). This means that the counteracting contributions from corresponding flux-charge phase differences are mismatch and subtracted with non-vanishing residue with a sign biasing the direction. However, for fluxes of opposite phase $\alpha_-=(2 n+1) \pi$ they add together and the magnitude instead depends on the ratio $(j_{0+}D_-+j_{0-}D_+)/D_+D_-$. Thus, switching the ionic diffusivity in this case may change the magnitude but does not affect the sign of the resulting unidirectional part and the net flow direction is preserved.

\begin{figure*}[t!]
	\centering
	\includegraphics[width=1\textwidth]{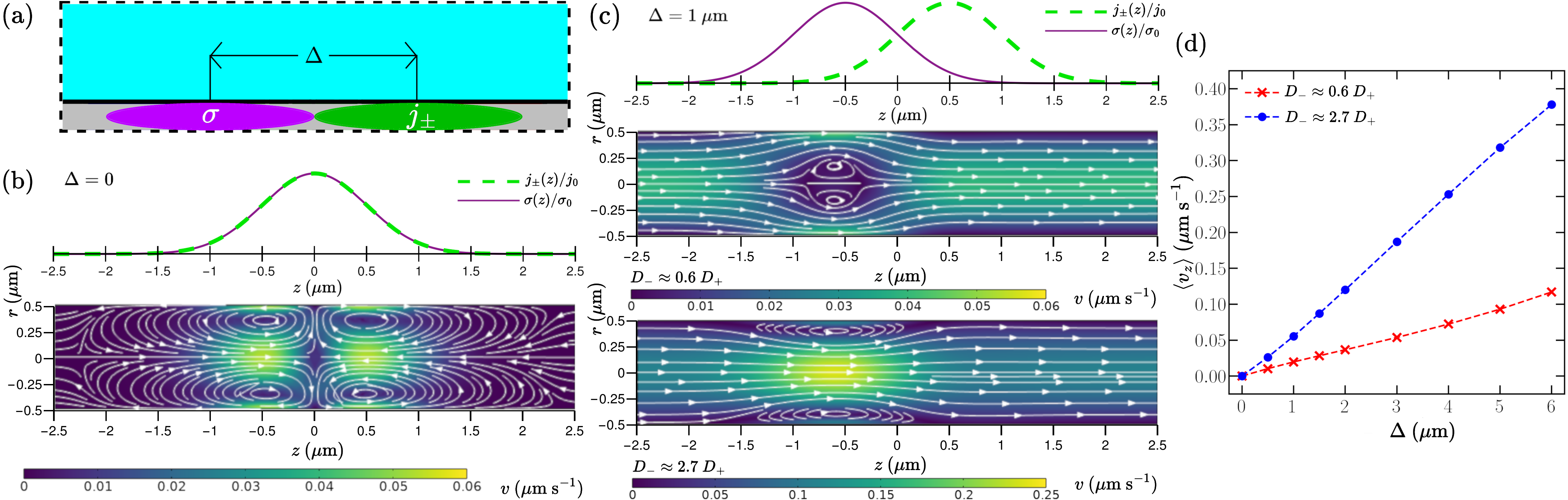}
	\caption{Enzyme powered flows generated by flux-charge patch separation and enhancement from diffusivity mismatch in microfluidic chamber. (a) Illustration of the zoomed-in view as in Fig.~\ref{fig1} of the interface between the capillary walls and the liquid solution inside showing a flux patch (green) and a charge patch (purple) separated by distance $\Delta$. The stream plots of steady state fluid velocity in (r,z) obtained numerically (using  {\it COMSOL}, Appendix~\ref{appenNum}) for (b) $\Delta=0$ and (c) $\Delta=1$ $\mu$m with different ion diffusivity mismatch. The corresponding Gaussian forms of the cationic (+) flux, the anionic (-) flux and the surface charge are shown above. The arrows of streamlines indicate the flow direction and the color map indicates the velocity magnitude. (d) Plot of average unidirectional velocity as a function of patch separation for different diffusivity mismatch. We use here bulk concentration $C_\infty=0.1$ mM, capillary radius $R=0.5$ $\mu$m, $j_0=5 \times 10^{-4}$ mM m s$^{-1}$, $\sigma_0=-1 \times 10^{-5}$ C m$^{-2}$, and Gaussian spread factor $\xi=2$ $\mu$m$^{-2}$.}
	\label{fig5}
\end{figure*}

Moreover, the switching of ionic diffusivity (that is changing from $D_+<D_-$ case to $D_+>D_-$ case, or vice versa) can also result in a change of state with a different flow structure depending on the flux and charge pattern configurations. For instance at $\beta=\pi/2$, when $\alpha_-=0$ the switching of diffusivity leads to complete flow reversal with a similar flow pattern but only flipped in direction (Fig.~\ref{fig4}(a)). Whereas for $\alpha_-=\pi/2$ the diffusivity switching results in a change of state with distinctive flow patterns without net flow reversal (Fig.~\ref{fig4}(b)). Meanwhile the averaged flow velocity for some arbitrary offset between the fluxes that is not of the same or opposite, like $\alpha_-=\pi/2$, the diffusivity switching can change its magnitude or direction depending on $\beta$ but at certain values both magnitude and direction can be left unchanged (Fig.~\ref{fig4}(c)). Additionally, to illustrate the change in averaged velocity landscape due to switching of cation-anion diffusivity, we provide the contour plot on $(\alpha_-, \beta)$ of the normalized averaged velocity for the two cases in Fig.~\ref{fig4}(d). 

In a mathematical sense, although our steady state Nernst-Planck equations in Eq.~(\ref{eqNP}) are independent of the diffusivity coefficients, the ionic flux boundary conditions in Eq.~(\ref{eqBC2}) however are not. Hence, the diffusivity of the cationic and anionic species also controls the relative flux amplitudes $\delta_\pm$ of Eq.~(\ref{deltas}), which the coefficients $h_{0,c,s}$ in Eq.~(\ref{eqh0})-(\ref{eqhS}) that determine the leading flow velocity rely upon. In a physical sense, difference in diffusivity of the cationic and anionic species creates a charge separation in the fluid at the steady state. In regions away from the surface where the charge separation persists, additional electric fields are induced in response to mitigate this separation and bring it close to neutrality in the bulk. Near the surface, this can incur additional mismatched ionic gradients. When diffusivity is switched, the charge separation reverses and the direction of the induced fields are hence flipped. The flow is affected by the overall contribution of these additional fields and gradients induced separation between the ionic species. Taken together, this effect of diffusivity mismatch can be an attribute of the additional gradients induced at the steady state as a consequence of difference in the effective flux magnitudes, which in turn affects the flow. This mismatch acts as an ion-specific parameter that controls the flow with the ability to optimize the speed as well as induce velocity reversal or change of flow state.

\section{Enzyme powered microfluidic generator} \label{secExample}

We provide here a concrete example of a microfluidic generator \cite{Sengupta2014,Song2024,DeCorato2020,Popescu2025} where an enzymatic reaction is applied to demonstrate the self-generated flows by the mechanism of flux-charge separation and diffusivity mismatch. Consider a microfluidic chamber, like the capillary-reservoir setup of Fig.~\ref{fig1}, where now a patch or region on the capillary wall is coated with an enzyme that catalyzes reactants which are mixed in the liquid solution and produces ionic species after the reaction. In particular, we look at two cases of urease-coated patches employed in previous works where the relevant end products are: (i) NH$^{+}_4$ and HCO$_3^{-}$ in Ref.~\cite{Song2024} and (ii) NH$^{+}_4$ and OH$^{-}$ in Ref.~\cite{DeCorato2020}. The key difference of interest between case (i) and (ii) is the ionic diffusivity mismatch that is $D_- \approx 0.6 D_+ $ and $D_- \approx 2.7 D_+ $, respectively \cite{Lide2005}. In conjunction with this enzyme patch, we also apply a patch or wall domain that induces a region of non-zero surface charge. This can be achieved as in previous works using silica \cite{DeCorato2020} or polymer coatings \cite{Stroock2000}. We focus here on flows due to a single pair of such flux-charge patches with a separation distance as illustrated in Fig.~\ref{fig5}(a). 
\subsection{Gaussian patch model}
We consider a finite sized axisymmetric system where a cylindrical capillary of diameter $1$ $\mu$m  ($R=0.5$ $\mu$m) and length $10$ $\mu$m connects two large electrolyte reservoirs (see Appendix~\ref{appenNum}). At the steady state, we assume wall patch distributions of ionic flux and charge density takes a Gaussian form given as follows 
\begin{equation}
    j_{\pm}(z)= j_0 e^{-\xi(z-\Delta/2)^2}  \quad \textrm{at}\quad  r=R
\end{equation}
\begin{equation}
    \sigma(z)= \sigma_0 e^{-\xi(z+\Delta/2)^2}  \quad \textrm{at}\quad  r=R
\end{equation}
where the Gaussian spread factor $\xi$ characterizes the patch size and the flux-charge separation distance $\Delta$ characterizes the displacement between sites of the peak densities. Based on the previous work \cite{DeCorato2020}, the magnitude of cationic and anionic flux settles to $j_0 \sim 10^{-4}$ mM m s$^{-1}$ (in-flux for urease) and the surface charge to  $\sigma_0 \sim 10^{-5}$ C m$^{-2}$ (negative for silica). These are micron sized patches corresponding to $\xi \sim 1$ $\mu$m$^{-2}$ and we vary $\Delta$ from $0$ up to $\sim 5$ $\mu$m (see top panels of Fig.~\ref{fig5}(b) and (c)).  
\subsection{Flows generated by flux-charge patch}
In panels (b)-(d) of Fig.~\ref{fig5}, we display the numerical results obtained for this patch model. When $\Delta=0$ there is complete overlap between the patch of flux and charge and leads to a circulatory flow at the site of peak density (Fig.~\ref{fig5}(b)). However, when they are displaced with a nonzero separation between their peak sites such as $\Delta=1$ $\mu$m, a unidirectional component develops and gives rise to localized vortices near at the site of peak charge (Fig.~\ref{fig5}(c)). These spots have velocity profiles representative of the intermediate state as shown in Fig.~\ref{fig2}(a) of $U\pm W$ corresponding to the different cases of diffusivity mismatch. We find that the magnitude of the directed flow increase with larger separation distance and is enhanced for the case with larger diffusivity mismatch (Fig.~\ref{fig5}(d)). We note that this net flow velocity can be further optimized by grafting a patterned sequence of multiple pairs of such flux-charge patches.

\section{Conclusions and outlook} \label{secConc}
In this paper we established a boundary-driven electrokinetic effect arising from active-charged patterns on confinement walls leading to directed electrolyte flow. Employing a continuum hydrodynamic description, we performed analytical approximations and validated with numerical finite element simulations. We show that modulation patterns of ionic fluxes and surface charge can self-generate a spectrum of flow states with a circulatory component, and notably, a unidirectional component that emerge when the flux and charge boundary patterns are offset by a flux-charge phase difference associated with the spatial mismatch between sites of peak flux and charge distributions. This can be achieved with just one ionic (cationic or anionic) flux active or with both fluxes active and exists even in the absence of externally applied fields or gradients in the bulk. We find selection rules with a rich dependence on the phase differences and amplitudes of fluxes and charge patterns lead the various flow states. When strong ionic flux constantly disrupts a non-uniform double layer, emergent lateral body forces become prominent driving fast flows. Varying key physical parameters such as the ionic diffusivity also affects flow fields and net velocity, resulting in velocity reversal, switching of the flow state and speed enhancement. Taken together, our work provides a theoretical framework for a boundary-driven mechanism based on non-uniform patterning of ionic activity and surface charges, that self-generates directed flows in confined environments. 

Cellular boundaries often involve active and charged biological structures. To shed light on the underlying physical mechanisms behind various cellular processes, it is important to understand the role of boundary features that can regulate fluid motion in micro- and nano-confinements \cite{Aggarwal2023,Aggarwal2023b,Kirkinis2022a}. At the same time, current developments in fluidic and iontronic technologies based on biological or soft matter systems, require control parameters that can functionalize the flow of confined electrolytes or ionic solutions. Our work here demonstrates a non-equilibrium steady state behavior of active-charged patterns inside capillaries that produce directed electrokinetic flows and gives quantitative insights on how such surface heterogeneties affect the state and magnitude of flow. Moreover, this work can be extended for systems of multiple active-charged channels or capillaries connected with each other, forming networks or circuits \cite{Henrique2024,Woodhouse2017}. For instance, in the circuit connections employed in fluidic or iontronic devices comprised of series, parallel or various combined configurations involving generalized network interactions of such capillary generators. Questions regarding the emerging physical principles governing active-charged networks can be further explored. In particular, those factors that determine fluid flow and ionic conductance for different network configurations, and tunable specifications that lead to optimal transport.

\begin{acknowledgments}
This work has been supported by the Department of Energy (DOE), Office of Basic Energy Sciences under Contract DE-FG02-08ER46539.
\end{acknowledgments}

\appendix

\appsection{Leading order equations} \label{appenLeq}
We give in this section the dimensionless forms of the reduced steady state governing equations with boundary conditions, and then write the leading order set of equations in the small-amplitude and long-wavelength approximations. The dimensionless transformations mentioned in Sec.~\ref{subsecPeturb} turns Eqs.~(\ref{eqP}) and~(\ref{eqNP}) into
\begin{equation} \label{eqDL1}
    \tilde{\nabla}^2 \tilde{\phi} = -  \tilde{\rho} \,,
\end{equation}
\begin{equation} \label{eqDL2}
    \tilde{\nabla}^2 \tilde{\rho} + \nabla (\tilde{s} \tilde{\nabla} \tilde{\phi})  = 0 \,,
\end{equation}
\begin{equation} \label{eqDL3}
    \tilde{\nabla}^2 \tilde{s} + \tilde{\nabla} (\tilde{\rho} \tilde{\nabla} \tilde{\phi})  = 0 \,,
\end{equation}
and the dimensionless boundary conditions at the surface become,
\begin{equation} \label{eqDLbc1}
    \partial_{\tilde{r}} \tilde{\phi} \ |_{\tilde{r}=\tilde{R}} = \delta_\sigma  \cos{(\tilde{k} \ \tilde{z}+\beta )} \,,
\end{equation}
\begin{equation} \label{eqDLbc2}
     (\partial_{\tilde{r}} \tilde{\rho} + \tilde{s} \partial_{\tilde{r}} \tilde{\phi} ) \ |_{\tilde{r}=\tilde{R}} = \delta_+  \cos{(\tilde{k} \tilde{z} + \alpha_+)} -\delta_- \cos{(\tilde{k} \ \tilde{z}+\alpha_- )} \,,
\end{equation}
\begin{equation} \label{eqDLbc3}
     (\partial_{\tilde{r}} \tilde{s} + \tilde{\rho} \partial_{\tilde{r}} \tilde{\phi} ) \ |_{\tilde{r}=\tilde{R}} = \delta_+   \cos{(\tilde{k} \ \tilde{z} + \alpha_+ )} +\delta_-  \cos{(\tilde{k} \ \tilde{z}+\alpha_- )} \,.
\end{equation}
By putting $\tilde{\nabla} \tilde{\phi} \approx \tilde{\nabla} \delta \tilde{\phi}$,  $\tilde{\rho} \approx \delta \tilde{\rho}$ and $\tilde{s} \approx 1+\delta \tilde{s}$ into Eqs.~(\ref{eqDL1}) and~(\ref{eqDL2}), we get the leading order set of equations,
\begin{equation}
    (\tilde{\nabla}^2 -1) \delta \tilde{\rho}  = 0 \,,
\end{equation}
\begin{equation}
    \tilde{\nabla}^2 \delta \tilde{\phi} = -  \delta \tilde{\rho} \,,
\end{equation}
that are constrained to the corresponding boundary conditions from Eqs.~(\ref{eqDLbc1}) and~(\ref{eqDLbc3}) which become,
\begin{equation}
\begin{split}
     \partial_{\tilde{r}} \delta \tilde{\rho} \ |_{\tilde{r}=\tilde{R}} = & \ \delta_+  \cos{(\tilde{k} \tilde{z} +\alpha_+)} -\delta_- \cos{(\tilde{k} \tilde{z}+\alpha_- )} \\
     & -\delta_\sigma  \cos{(\tilde{k} \tilde{z}+\beta )} \,,
\end{split}     
\end{equation}
\begin{equation}
    \partial_{\tilde{r}}  \delta \tilde{\phi} \ |_{\tilde{r}=\tilde{R}} = \delta_\sigma  \cos{(\tilde{k} \ \tilde{z}+\beta )} \,,
\end{equation}
and can be solved by separation of variable method \cite{Arfken2012}. This gives the analytical expressions given in Eqs~(\ref{solrho}) and (\ref{solphi}) in the small-amplitude approximations. Then to obtain the expression for the body force the respective $\tilde{r},\tilde{z}$ derivatives of Eqs~(\ref{solrho}) and (\ref{solphi}) are taken. The resulting curl of this body force is now plugged into the right hand side of Eq.~(\ref{eqVort}) giving 
\begin{equation} \label{eqVortApp}
    \tilde{\mathcal{D}}^4 \tilde{\psi} \approx \epsilon_{ij} G_{i,j}(\tilde{r}) [h^{(i,j)}_{0}+ h^{(i,j)}_{c} \cos{(2 \tilde{k} \tilde{z})} + h^{(i,j)}_{s} \sin{(2 \tilde{k} \tilde{z})}]  \,.
\end{equation}
where $\epsilon_{ij}$ is the Levi-Civita symbol using Einstein summation notation over $i,j \in \{1,2\}$. In the long wavelength approximation, the terms up to $\mathcal{O}(\tilde{k}^{-1})$ become,
\begin{equation} \label{eqVortAppG12}
    G_{1,2}(\tilde{r}) = \frac{\tilde{r} \ I_1(\sqrt{1+\tilde{k}^2} \tilde{r}) I_0(\tilde{k}\tilde{r})}{2 \ I_1(\sqrt{1+\tilde{k}^2}\tilde{R})I_1(\tilde{k}\tilde{R})}    \approx  \frac{\tilde{r}}{\tilde{k} \tilde{R}} \frac{I_1(\tilde{r})}{I_1(\tilde{R})} \,,
\end{equation}
\begin{equation} \label{eqVortAppG21}
    G_{2,1}(\tilde{r}) = \frac{\tilde{k}\tilde{r}}{2 \sqrt{1+\tilde{k}^2}} \frac{I_0(\sqrt{1+\tilde{k}^2} \tilde{r}) I_1(\tilde{k}\tilde{r})}{I_1(\sqrt{1+\tilde{k}^2}\tilde{R})I_1(\tilde{k}\tilde{R})}   \approx 0 \,.
\end{equation}
This further reduces Eq.~(\ref{eqVortApp}) to
\begin{equation} \label{eqVortAppRed}
    \tilde{\mathcal{D}}^4 \tilde{\psi} \approx \frac{\tilde{r} \ I_1(\tilde{r})}{\tilde{k} \tilde{R} \ I_1(\tilde{R})} [h_{0}+ h_{c} \cos{(2 \tilde{k} \tilde{z})} + h_{s} \sin{(2 \tilde{k} \tilde{z})}]  \,.
\end{equation}
which is then solved for stream function accompanied with the no-slip conditions Eq.~(\ref{eqBCns}) \cite{Happel1983, Anderson1985}. We note that in the leading order the equations involving perturbed field $\delta \tilde{s}$ are decoupled and hence not necessary to obtain for the purposes of the leading order stream function and flow velocity.

\begin{figure}[t!]
   \centering
    \includegraphics[width=0.95\columnwidth]{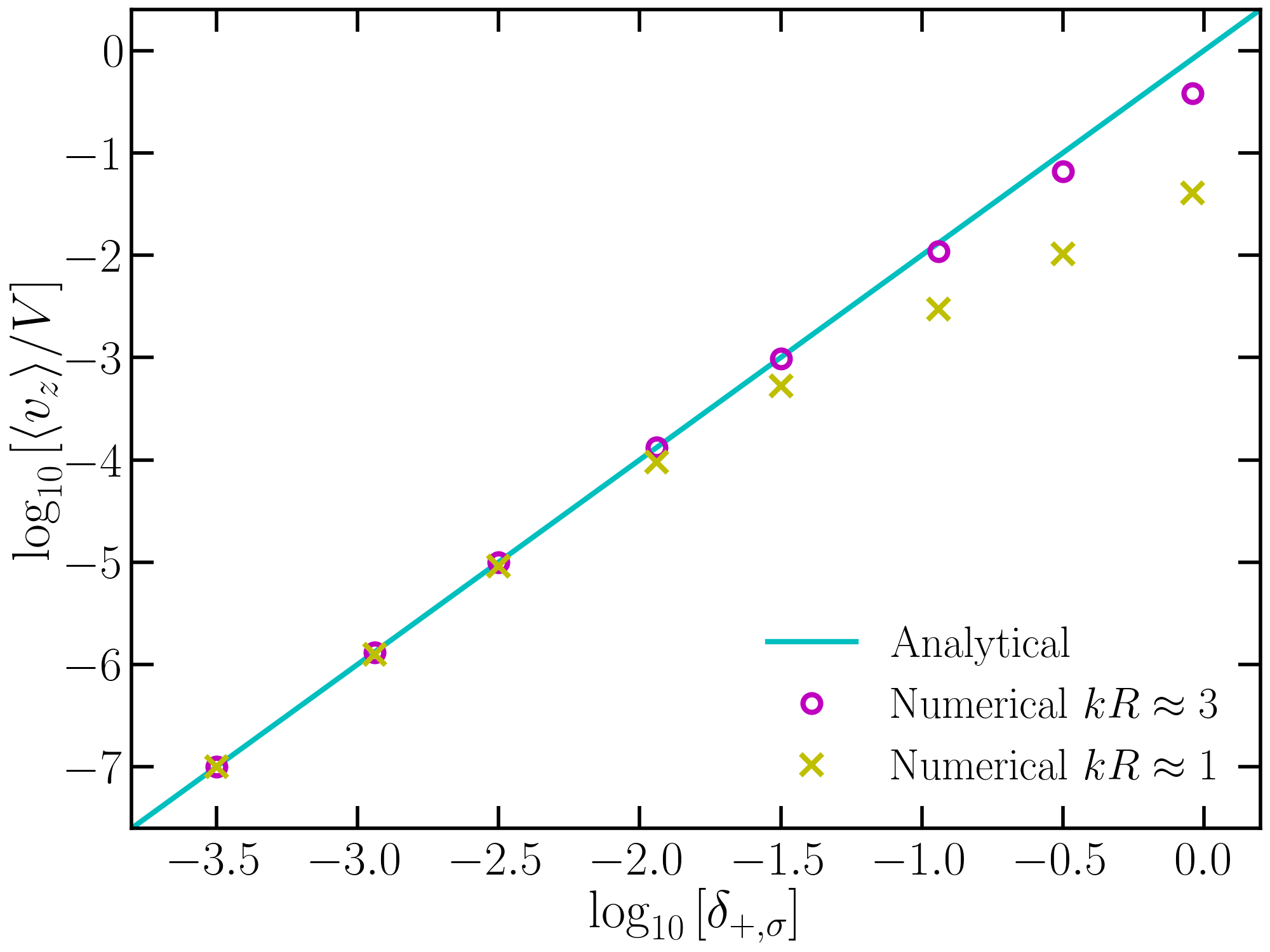}
    \caption{Log-log plot of average unidirectional velocity as a function of relative flux-charge amplitudes $\delta_{+,\sigma}$ at different $kR$. The analytical expression Eq.~(\ref{eqLog}) is compared with the numerical results from \emph{COMSOL}, and $V$ is given by Eq.~(\ref{eqVmax}). We use here $C_\infty=1$ mM  and $R=0.5$ $\mu$m.}
    \label{fig6}
\end{figure}

\appsection{Small-amplitude and long-wavelength approximation} \label{appenApprox}
In this section, we look at the limits for which the small-amplitude and long-wavelength approximations hold. The ionic fluxes and surface charge are taken to be small such that the perturbed concentration fields are $\delta c_\pm \ll C_\infty/2$ or $\delta \tilde{\rho} \ll 1$ which gives in the long wavelength limit, 
\begin{equation}
    \frac{I_0(\sqrt{1+ \tilde{k}^2}\tilde{R})}{\sqrt{1+ \tilde{k}^2} \ I_1(\sqrt{1+ \tilde{k}^2}\tilde{R})} \delta_{\pm,\sigma} \approx \delta_{\pm,\sigma} \ll 1
\end{equation}
Along with this, the magnitude of the electric field is also considered small, that is $|\tilde{\nabla} \delta \tilde{\phi}| \ll 1$. The $r$-component ($\partial_{\tilde{r}} \tilde{\phi}$) gives a similar condition $\delta_{\pm,\sigma} \ll 1$ and the $z$-component  ($\partial_{\tilde{z}} \tilde{\phi}$) we have again $\delta_\sigma \ll 1$ but also
\begin{equation}
 \left| \frac{I_0(\tilde{k}\tilde{R})}{I_1(\tilde{k} \tilde{R})}- \frac{\tilde{k}\ I_0(\sqrt{1+ \tilde{k}^2}\tilde{R})}{\sqrt{1+ \tilde{k}^2} \ I_1(\sqrt{1+ \tilde{k}^2}\tilde{R})} \right| \ \delta_{\pm} \approx \frac{2 \delta_{\pm,\sigma}}{\tilde{k} \tilde{R}} \ll 1
\end{equation}
This implies restrictions on wavelength and radius such that $\tilde{k} \tilde{R} > 2 \delta_{\pm} $. For example, if $l \approx 20R$ so that $kR= 2\pi R/l \approx 0.3$ then $\delta_\pm \lesssim 0.15$, or conversely if $\delta_\pm \approx 0.1$, then $l \lesssim 30 R$.

We compare the analytical and numerical average velocity for the case $\delta_+ \approx \delta_\sigma$ where the analytical expression of Eq.~(\ref{eqVdel}) written in log-log form is
\begin{equation} \label{eqLog}
   \log_{10}[ \langle v_z \rangle/V] \approx 2 \log_{10}[\delta_{+,\sigma}]
\end{equation}
where $V$ is given by Eq.~(\ref{eqVmax}). Fig.~\ref{fig6} shows the log-log plot comparing the analytical and numerical values, where we find an excellent matching for $\delta_{+,\sigma} \lesssim 0.1$ with $kR \gtrsim 2$.

\begin{figure*}[t!]
    \centering
    \includegraphics[width=0.9\textwidth]{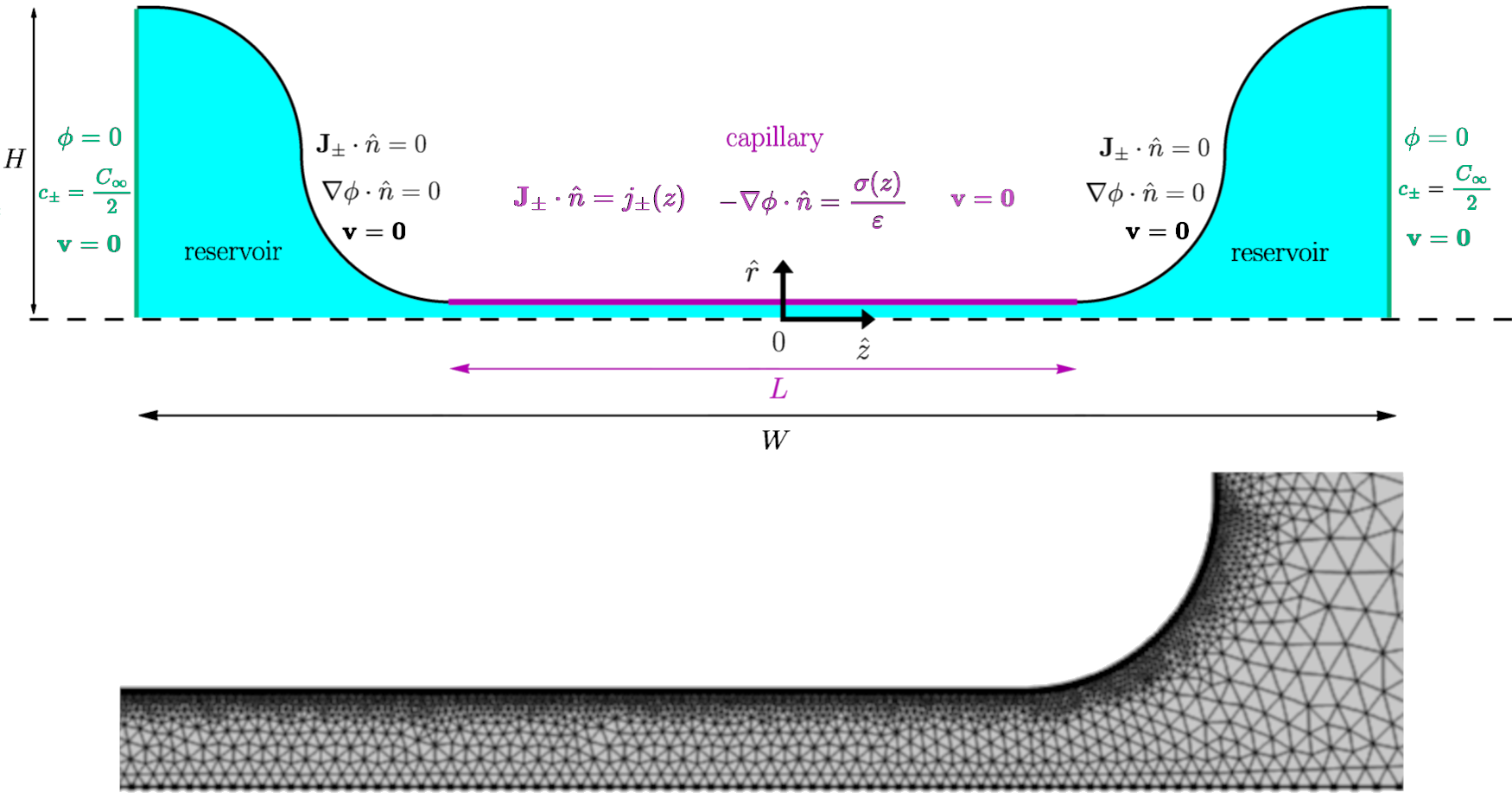}
    \caption{Top panel: Schematic cross-section of the set-up employed in the numerical scheme. A cylindrical capillary of radius $R$ and length $L$ connecting two identical reservoirs make up the capillary-reservoir system of size $W$ by $H$. Shown here are the boundary conditions applied at different regions of the walls enclosing the electrolyte. At the capillary surface the conditions are given by Eqs.~(\ref{eqBC1}) and~(\ref{eqBC2}), while connecting regions are set to satisfy the zero normal electric field and zero normal flux conditions. We impose no-slip and no-penetration conditions for the velocity on the walls. The far reservoir ends have conditions of zero electric potential and concentrations set to its bulk value and zero velocity. Bottom panel: Mesh grid pattern (zoomed in view of right hand section) implemented in {\it COMSOL Multiphysics} software \cite{comsol} for numerically convergent solutions. The mesh sizes are adjusted (for $\lambda=0.1 R$) with a finer mesh of $2 \times 10^{-4} R$ near the wall and sparser mesh of $2 \times 10^{-2} R$ away from it in the capillary domain, and then to larger sizes up to $0.1 H$ in the reservoirs. }
\label{fig7}  
\end{figure*}

\appsection{Streamfunction coefficients} \label{appenCoeff}
The streamfunction \rr{eqStream} and resulting velocity fields \rr{eqVr} and \rr{eqVz} depend on the functions
\begin{equation} \label{eqStream0}
    \Psi_0(\tilde{r}) = A_0 + B_0 \tilde{r}^2 + \frac{\tilde{r} I_1(\tilde{r})}{\tilde{k} \tilde{R} \ I_1(\tilde{R})}   \,, 
\end{equation}
and
\begin{equation} \label{eqStream1}
    \Psi_1(\tilde{r}) = A_1 \tilde{r}^2 I_0(2 \tilde{k}\tilde{r}) + B_1 \tilde{r} I_1(2 \tilde{k} \tilde{r}) + \frac{\tilde{r} I_1(\tilde{r})}{\tilde{k} \tilde{R} \ I_1(\tilde{R})}  \,,
\end{equation}
where the coefficients $A_0,B_0,A_1,B_1$ are functions of $\tilde{k}$ and $\tilde{R}$
\begin{equation}
    A_0 =  \frac{1}{\tilde{k}} \left( \frac{ \tilde{R} I_0(\tilde{R})}{2  \ I_1(\tilde{R})} -1 \right) + \mathcal{O}(\tilde{k}^0) \,, 
\end{equation}
\begin{equation}
    B_0 = - \frac{I_0(\tilde{R})}{2 \tilde{k} \tilde{R} \ I_1(\tilde{R})}  + \mathcal{O}(\tilde{k}^0) \,,
\end{equation}
\begin{widetext}
\begin{equation}
    A_1  = \frac{1}{\tilde{k}^3 \tilde{R}^3}  \left( \frac{ I_0(\tilde{R})}{I_1(\tilde{R})} - \frac{2}{\tilde{R}} \right) + \frac{1}{\tilde{k} \tilde{R}}  \left[  \frac{8}{\tilde{R}^2}\left( \frac{ I_0(\tilde{R})}{I_1(\tilde{R})} - \frac{2}{\tilde{R}} \right)- \frac{ I_0(\tilde{R})}{6 \ I_1(\tilde{R})} - \frac{2}{3 \tilde{R}} \right] + \mathcal{O}(\tilde{k}^0) \,,
\end{equation}
\begin{equation}
    B_1  = \frac{1}{\tilde{k}^4 \tilde{R}^4}  \left(2 - \frac{ \tilde{R} I_0(\tilde{R})}{I_1(\tilde{R})} \right) + \frac{1}{\tilde{k}^2 \tilde{R}^2}  \left[ \frac{8}{3} -  \frac{ 
\tilde{R} I_0(\tilde{R})}{3 \ I_1(\tilde{R})} - \frac{8}{\tilde{R}}\left( \frac{ I_0(\tilde{R})}{I_1(\tilde{R})} - \frac{2}{\tilde{R}} \right)  \right] + \mathcal{O}(\tilde{k}^0) \,.
\end{equation}
\end{widetext}

\appsection{Numerical finite element simulations} \label{appenNum}
In this section we supplement the details about the scheme employed for the numerical solution of the steady state PNPNS system and the averaged velocity. We perform finite element simulations using the {\it COMSOL Multiphysics} software \cite{comsol}. As in the main text, we use the time-independent PNPNS equations reduced at the low P\'eclet and low Reynolds number regime, leaving the set of Eqs.~(\ref{eqP})-(\ref{eqS2}). This coupled system is solved in a cylindrical geometry for the set-up displayed in Fig.~\ref{fig7}. Differences between the flow velocity obtained numerically and the analytically, are attributed to the small-amplitude and long-wavelength approximations employed to obtain the analytical solutions and the finite length set-up implemented for the numerical simulation. However, despite these differences the results agree well as shown in Fig.~\ref{fig3} of the main text and in Fig.~\ref{fig5} of Appendix~\ref{appenApprox}. 

The set-up of this numerical scheme entails an axisymmetric cylindrical geometry where a capillary of radius $R$ and length $L$ connects two identical reservoirs as shown in Fig.~\ref{fig7}, similar to configurations employed in previous works. The boundary conditions at the capillary surface are the same as those employed in the main text with Eqs.~(\ref{eqBC1}) and~(\ref{eqBC2}). The regions connecting the reservoirs to the capillary are set to satisfy the condition of zero normal electric field $\nabla \phi \cdot \hat{n}=0$ and also the zero normal flux $\mathbf{J}_\pm \cdot \hat{n}=0$. Finally, we impose no-slip and no-penetration boundary conditions for the velocity on all the bounding walls. Finally, at the far edges of the capillary-reservoir system of size $W$ by $H$, we fix the concentration to the bulk value $c_\pm= C_\infty/2$ and set electric potential to zero $\phi=0$ with the fluid velocity to zero $\mathbf{v}=\mathbf{0}$. In {\it COMSOL} we conduct a stationary (time-independent) study of this set-up by implementing an interface coupling between the {\it Electrostatics} ({\it es}) and the {\it Transport of Diluted Species} ({\it tds}) interfaces provided by the {\it Chemical Reaction Engineering} module, in conjunction with the {\it Creeping Flow} ({\it spf}) interface provided by the {\it Computational Fluid Dynamics} module \cite{comsol}. This scheme which amounts to computing the numerical solutions of the coupled system Eqs. (\ref{eqP})-(\ref{eqS2}) with the described boundary conditions is executed for a given set of input parameters. 

The system sizes are set according to the capillary radius and wavelength, for instance in Fig.~\ref{fig3} we use $L=180R$, $W=200R$ and $H=10R$. While for the patch model in Fig.~\ref{fig5} we use $L=10$ $\mu$m, $W=30$ $\mu$m and $H=10$ $\mu$m. Numerically convergent solutions require appropriate sizes of mesh grid elements in accordance to the system length scales over which our solutions are expected to vary. These scales rely here on parameters such as the Debye length and the flux-charge wavelengths. For instance, larger variations of the electric potential typically occur within the Debye layer near the surface rather than in the bulk far away from it. Thus, near the charged capillary walls a finer mesh is needed with sizes smaller than $\lambda = 0.1 R$ here. In this respect, we sequentially adjust our mesh sizes inside our capillary domain with a finer mesh of $2 \times 10^{-4} R$ near the wall and sparser mesh of $2 \times 10^{-2} R$ away from it, and then to larger sizes of up to $0.1 H$ in the reservoirs (see Fig.~\ref{fig7}). As a result, with these settings {\it COMSOL} provides convergent solutions of the electric potential, ion concentrations and fluid velocity for the set-up and parameters employed here.

\bibliographystyle{apsrev4-2}
\bibliography{reference}

\end{document}